\documentclass[english,prd,twocolumn,superscriptaddress,showpacs,preprintnumbers]{revtex4-2}

\usepackage{graphicx}
\usepackage{multirow}
\usepackage{mathrsfs}
\usepackage{amsmath}
\usepackage{float}
\usepackage{xcolor}
\usepackage[colorlinks,linkcolor=blue,citecolor=blue,urlcolor=blue]{hyperref}

\begin{document}

\title{Effective Field Theory Analysis of CDMSlite Run 2 Data}

\author{M.F.~Albakry} \affiliation{Department of Physics \& Astronomy, University of British Columbia, Vancouver, BC V6T 1Z1, Canada}\affiliation{TRIUMF, Vancouver, BC V6T 2A3, Canada}
\author{I.~Alkhatib} \affiliation{Department of Physics, University of Toronto, Toronto, ON M5S 1A7, Canada}
\author{D.W.P.~Amaral} \affiliation{Department of Physics, Durham University, Durham DH1 3LE, UK}
\author{T.~Aralis} \affiliation{Division of Physics, Mathematics, \& Astronomy, California Institute of Technology, Pasadena, CA 91125, USA}
\author{T.~Aramaki} \affiliation{Department of Physics, Northeastern University, 360 Huntington Avenue, Boston, MA 02115, USA}
\author{I.J.~Arnquist} \affiliation{Pacific Northwest National Laboratory, Richland, WA 99352, USA}
\author{I.~Ataee~Langroudy} \affiliation{Department of Physics and Astronomy, and the Mitchell Institute for Fundamental Physics and Astronomy, Texas A\&M University, College Station, TX 77843, USA}
\author{E.~Azadbakht} \affiliation{Department of Physics and Astronomy, and the Mitchell Institute for Fundamental Physics and Astronomy, Texas A\&M University, College Station, TX 77843, USA}
\author{S.~Banik} \affiliation{School of Physical Sciences, National Institute of Science Education and Research, HBNI, Jatni - 752050, India}
\author{C.~Bathurst} \affiliation{Department of Physics, University of Florida, Gainesville, FL 32611, USA}
% ...
\collaboration{SuperCDMS Collaboration}

\begin{abstract}
CDMSlite Run 2 was a search for weakly interacting massive particles (WIMPs) with a cryogenic 600\,g Ge detector operated in a high-voltage mode to optimize sensitivity to WIMPs of relatively low mass from 2 -- 20\,GeV/$c^2$.
In this article, we present an effective field theory (EFT) analysis of the CDMSlite Run 2 data using an extended energy range and a comprehensive treatment of the expected background.
A binned likelihood Bayesian analysis was performed on the recoil energy data, taking into account the parameters of the EFT interactions and optimizing the data selection with respect to the dominant background components.
Energy regions within 5$\sigma$ of known activation peaks were removed from the analysis.  
The Bayesian evidences resulting from the different operator hypotheses show that the CDMSlite Run 2 data are consistent with the background-only models and do not allow for a signal interpretation assuming any additional EFT interaction.
Consequently, upper limits on the WIMP mass and coupling-coefficient amplitudes and phases are presented for each EFT operator.
These limits improve previous CDMSlite Run 2 bounds for WIMP masses above 5\,GeV/$c^2$.
\end{abstract}

\maketitle


\section{Introduction}

Dark matter (DM), first hypothesized by astrophysicists in the 1920s and 1930s to explain gravitational effects in galactic dynamics \cite{Kapteyn:1922zz,Zwicky:1933gu,Oort:1940eek}, makes up about 85\% of matter in the Universe \cite{Planck2018}.
Recent astrophysical evidence also supports the existence of DM \cite{Ade:2015xua,Planck2018,Clowe:2003tk,Aubourg:2014yra}.
Other than gravitational interactions, DM may also interact with baryonic matter with a cross section on the scale of the weak force \cite{Jedamzik:2009uy,Jungman:1995df,Cirelli:2005uq}, which motivates the direct search for DM using experiments sensitive to nuclear scattering from a variety of targets.
Dark matter particles of this type are known as weakly interacting massive particles (WIMPs).
In the galactic halo, WIMPs would be moving at typical speeds of approximately 250\,km/s.
Their elastic scattering off nuclei in an underground detector would deposit a small amount of energy, in the range of 0.1 to 100\,keV for a WIMP mass above 1\,GeV/$c^2$~\cite{Lewin:1995rx}.

A general description of the WIMP-nucleon interaction has been derived using an effective field theory (EFT) approach \cite{Fan:2010gt,Fitzpatrick:2012ix,Fitzpatrick:2012ib,Anand:2013yka}.
This formalism extends the model-driven conventional technique -- assuming the simplest spin- and isospin-independent interactions -- by considering all relevant couplings in the non-relativistic limit \cite{Catena:2014uqa,DelNobile:2013sia}.
The addition of angular-momentum-dependent and spin-and-angular-momentum-dependent couplings means that the EFT approach includes interaction operators that are also dependent on momentum transfer and the initial velocities, which changes the shape of the nuclear recoil energy spectrum for some operators \cite{Fitzpatrick:2012ix}.
The experimental response to the different EFT contributions is also very sensitive to the properties of the nuclear target, providing an excellent motivation for the search with different experimental setups.
Limits on EFT operators have been obtained by SuperCDMS \cite{SuperCDMS:2015lcz}, PandaX \cite{PandaX-II:2018woa}, XENON \cite{ XENON:2017fdd}, DEAP \cite{DEAP:2020iwi}, and LUX \cite{ LUX:2020oan, LUX:2021ksq}.

The Super Cryogenic Dark Matter Search (SuperCDMS) experiment uses cryogenically cooled Ge and Si detectors to measure nuclear recoils via ionization and phonon signals \cite{Agnese:2013rvf,Agnese:2015ywx,Agnese:2016cpb}.
One mode of operating SuperCDMS detectors, known as the CDMS low ionization threshold experiment (CDMSlite), increases the amplification voltage across the crystal in order to detect lower energy recoils \cite{Agnese:2013jaa,Agnese:2015nto,Agnese:2017jvy}.
This allows for a better reach to lower WIMP masses as well as discrimination between the nuclear recoil energy spectra of different EFT operators.
Operating the detector in this mode trades discrimination between electron and nuclear recoils for lower energy thresholds.
For the current WIMP search, electron recoils are a background.

In this article, data from CDMSlite Run 2 are analyzed within the EFT framework, including higher recoil energies than previously published by the SuperCDMS collaboration \cite{Agnese:2015nto,Agnese:2017jvy}.
The article is organized as follows.
The CDMSlite setup and the background modeling for the Run 2 data are described in Sec.~\ref{sec:CDMSlite}.
In Sec.~\ref{sec:MLE} the EFT framework and the Bayesian likelihood analysis are described.
This analysis technique is applied in Sec.~\ref{sec:compare} to calculate an upper limit on the WIMP mass and coupling, comparing the result to the previously published CDMSlite Run 2 limit.
In Sec.~\ref{sec:evidence} the Bayesian evidences for all models considered are presented.
The resulting posterior distributions are discussed in Sec.~\ref{sec:likelihoods} for a two-parameter background-only model (Sec.~\ref{sub:bkg_only}) and for each EFT operator (Sec.~\ref{sub:single_op}).
Finally our conclusions are presented in Sec.~\ref{sec:conclusions}.
Appendix~\ref{app:priors} contains a discussion of several Bayesian prior probability options.

\section{CDMSlite Run 2} \label{sec:CDMSlite}

In a SuperCDMS detector, the energy deposited in a nuclear recoil produces prompt phonons and excites charge carriers into the conduction band.
The latter are then drifted to the crystal surfaces by applying a voltage difference across the crystal.
During the drift of the charge carriers, secondary Neganov-Trofimov-Luke (NTL) phonons are released \cite{Luke:1988gwe,Neganov:1985khw}.
In the conventional operation mode, when the voltage across the crystal is low ($\sim 4$\,V), electron recoils can be discriminated from nuclear recoils by comparing the strengths of the phonon and ionization signals, with excellent separation above $\sim 8$\,keV recoil energy \cite{Agnese:2013ixa}.

In the case of higher voltage bias across the crystal ($\sim 70$\,V), the production of secondary NTL phonons is amplified linearly with respect to the voltage bias.
Moreover, by applying a bias voltage the signal due to NTL phonons can be significantly increased without substantially increasing the phonon readout noise, thereby reducing the experimental threshold to focus on low-mass WIMP detection.
In this high-voltage (HV) mode, the ionization signal is not measured; therefore, in contrast to the low-voltage case, the nuclear and electron recoil events cannot be distinguished.
As a consequence, the background for HV data sets includes all electron recoil backgrounds, with the dominant contributions including the spectral lines due to various activation isotopes and the continuous backgrounds due to tritium contamination and Compton scattering.

``CDMSlite'' refers to operating the SuperCDMS detectors in this HV mode  \cite{Agnese:2013jaa,Agnese:2015nto,Agnese:2017jvy}.
\mbox{CDMSlite} Run 2 was performed at the Soudan Underground Laboratory in northern Minnesota in 2014, using a single $\sim 600$\,g Ge crystal.
Results from this run were previously published up to 20\,keV$_\text{ee}$ (electron equivalent energy) or $\sim60$\,keV$_\text{nr}$ (nuclear recoil energy) for the conventional spin-independent \cite{Agnese:2015nto} and spin-dependent cross sections \cite{Agnese:2017jvy}. 

In order to perform a binned likelihood analysis of the CDMSlite Run 2 data, the total exposure (WIMP detection efficiency\,$\times$\,detector mass\,$\times$\,live time), background models, and binned recoil energy spectrum must be defined.
The data spectrum contains all events passing the standard data-quality and fiducial cuts \cite{Agnese:2017jvy} from the experimental threshold of 0.3\,keV$_\text{nr}$ up to 60\,keV$_\text{nr}$ binned into 300 energy bins ($\Delta E \approx 0.2$\,keV$_\text{nr}$).
Because energy regions with high backgrounds do not contribute much to the overall experimental sensitivity, several regions around known activation peaks get excluded.
This approach also simplifies the background modeling.

\begin{figure}[t]
   \centering
   \includegraphics[width=1.02\columnwidth]{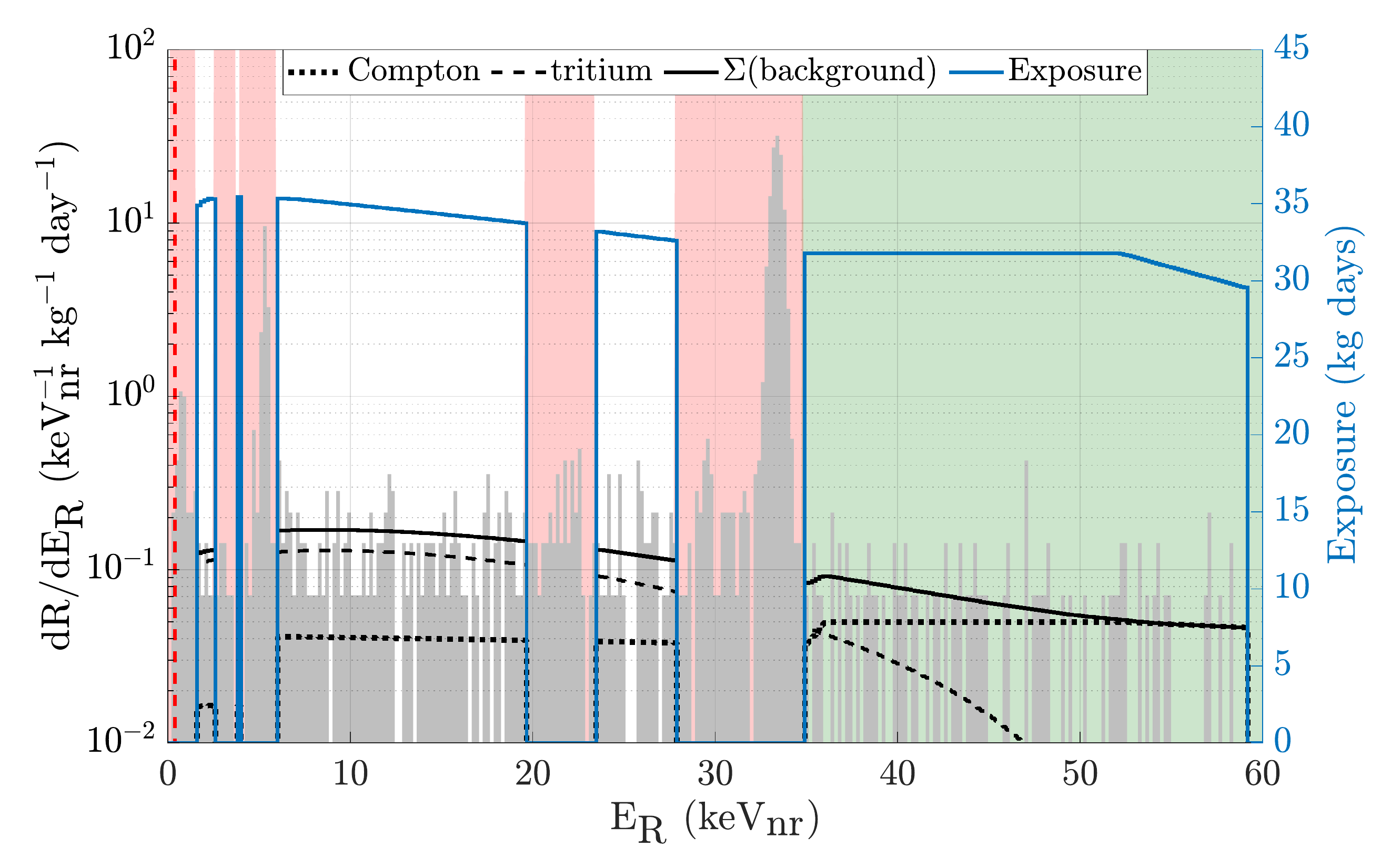}
   \caption{CDMSlite Run 2 observed differential recoil energy spectrum (light gray) and exposure (blue solid curves) with regions removed from the ROI (red vertical strips) and remaining background sources (dotted black = Compton, dashed black = tritium, solid black = total background). The vertical red dashed line indicates the experimental threshold of 0.3\,keV$_\text{nr}$. The green shaded region above 35\,keV$_\text{nr}$ designates the region which is dominated by Compton scattering and tritium contamination. The amplitudes of the background models shown here were chosen to show the shapes of the spectra and do not represent the best fits to the data.}
   \label{fig:CDMSlite_defined}
\end{figure}

The excluded regions are defined as within $5\sigma$ of the K-shell, L-shell, and M-shell peaks for $^{71}$Ge, $^{68}$Ge, $^{68}$Ga, $^{65}$Zn, $^{55}$Fe, and $^{54}$Mn, as detailed in Tab.~\ref{tab:activation_peaks}.
These isotopes were chosen because hints of their characteristic peaks are visible in the CDMSlite Run 2 spectrum.
This removes the lowest-energy events (labeled as Region 1 in Tab.~\ref{tab:activation_peaks}), effectively increasing the energy threshold to 1.6\,keV$_\text{nr}$.
The data spectrum, region of interest (ROI), as well as exposure and background models in the ROI are shown in Fig.~\ref{fig:CDMSlite_defined}.
For details on the efficiency and data spectrum see Ref.~\cite{Agnese:2019app}.

The backgrounds included in the analysis are the continuous recoil energy spectra due to tritium decay and Compton scattering.
Additional backgrounds from contamination and additional neutron activation peaks exist in the CDMSlite data set; however, the peaks from these contributions are below the sensitivity of the experiment and not clearly visible in the data set, so models for these backgrounds were not included in this analysis.
Background due to surface contamination of $^{210}$Pb is also not included; however, estimates
based on the observed $\alpha$-decay event rates in the detector suggest that it contributes not more than about 8\% to the continuous low-energy spectrum \cite{Agnese:2019app}.
Based on these estimates only $\sim 20$ $^{210}$Pb events are expected in the $0-60$\,keV$_\text{nr}$ energy range, which is much smaller than the Compton or tritium contributions (each estimated to be of the order of 200 events, as we will show in Sec.~\ref{sub:bkg_only}).

\begin{table}[ht!]
\caption{Energy ranges of the excluded regions, which were determined at 5$\sigma$ of the activation peaks from $^{71}$Ge, $^{68}$Ge, $^{68}$Ga, $^{65}$Zn, $^{55}$Fe, and $^{54}$Mn \cite{Lederer1968}. The rightmost column contains the peak energy and the observed resolution ($\sigma$).}
\label{tab:activation_peaks}
\begin{ruledtabular}
\begin{tabular}{cccc}
Region & Energy Range & Source & Peak Energy \\
 & (keV$_\text{nr}$) & (Shell) & (keV$_\text{nr}$) \\
\colrule
\multirow{6}{*}{1} & \multirow{6}{*}{0 -- 1.6} & $^{54}$Mn (M) & $0.40\pm0.06$ \\
& & $^{55}$Fe (M) & $0.44\pm0.06$ \\
& & $^{65}$Zn (M) & $0.63\pm0.06$ \\
& & $^{68}$Ga (M) & $0.71\pm0.06$ \\
& & $^{68}$Ge (M) & $0.80\pm0.08$ \\
& & $^{71}$Ge (M) & $0.80\pm0.08$ \\
\colrule
\multirow{2}{*}{2} & \multirow{2}{*}{2.4 -- 3.8} & $^{54}$Mn (L) & $3.01\pm0.09$ \\
& & $^{55}$Fe (L) & $3.30\pm0.09$ \\
\colrule
\multirow{5}{*}{3} & \multirow{5}{*}{4.0 -- 6.0} & $^{65}$Zn (L) & $4.5\pm0.1$ \\
& & $^{71}$Ge (L2) & $4.7\pm0.1$ \\
& & $^{68}$Ga (L) & $4.9\pm0.1$ \\
& & $^{68}$Ge (L) & $5.3\pm0.1$ \\
& & $^{71}$Ge (L) & $5.3\pm0.1$ \\
\colrule
\multirow{2}{*}{4} & \multirow{2}{*}{19.4 -- 24.4} & $^{54}$Mn (K) & $20.6\pm0.2$ \\
& & $^{55}$Fe (K) & $23.3\pm0.2$ \\
\colrule
\multirow{4}{*}{5} & \multirow{4}{*}{27.4 -- 35.0} & $^{65}$Zn (K) & $29.1\pm0.3$ \\
& & $^{68}$Ga (K) & $31.4\pm0.3$ \\
& & $^{68}$Ge (K) & $33.4\pm0.3$ \\
& & $^{71}$Ge (K) & $33.4\pm0.3$ \\
\end{tabular}
\end{ruledtabular}
\end{table}

The observed energy resolutions ($\sigma$) of the K, L, and M activation peaks, as well as of randomly triggered zero-energy events, were used to obtain the following best-fit relationship:
\begin{equation}
\sigma(E_{\text{ee}}) = \sqrt{\sigma_0^2+\varepsilon F E_{\text{ee}} +(A E_{\text{ee}})^2},
\end{equation}
where $\sigma_0=(9.26\pm0.11)$\,eV$_\text{ee}$, $F$ corresponds to the Fano factor contribution with $\varepsilon F=(0.64\pm0.11)$\,eV$_\text{ee}$, $\varepsilon=3$\,eV is the average energy needed to create a single electron-hole pair in Ge, $E_\text{ee}$ is the electron equivalent energy and \mbox{$A=(5.68\pm0.94)\times10^{-3}$} \cite{Aralis:2019nfa}.
In order to compare with the expected EFT rates, all energies must be converted to the nuclear recoil energy scale, denoted as $E_\text{R}$ in the following.
$E_\text{ee}$ is related to $E_\text{R}$ by
\begin{equation}
E_\text{ee} =  E_\text{R}\times\frac{(1+Y(E_\text{R})\times e\cdot V/\varepsilon)}{1+e\cdot V/\varepsilon},
\end{equation}
where $e$ is the electron charge, $V=70$\,V is the voltage across the CDMSlite detector and $Y(E_\text{R})$ is the ionization yield of a nuclear recoil \cite{Agnese:2017jvy} with $E_\text{R}$ in keV$_\text{nr}$.
We use the Lindhard yield model,
\begin{equation}
Y = \frac{k\times g}{1+k\times g},
\end{equation}
and 
\begin{equation}
g = 3x^{0.15}+7x^{0.6}+x
\end{equation}
with $x = 11.5\text{\,keV}^{-1}\, E_\text{R}\,Z^{-7/3}$ \cite{Lindhard1963a,Lindhard1963,Lindhard1968}.
For Ge, $k=0.157$ and $Z=32$.
For energies at and above our effective analysis energy threshold, the Lindhard model is a suitable choice to model the nuclear yield \cite{Barker:2012ap}.

The dominant background source outside the excluded regions is due to Compton scattering and tritium contamination.
These backgrounds were modeled as follows.
In order for an electron recoil to occur, the electron must have at least the discrete binding energy of the shell it came from.
Therefore, the low energy Compton scattering spectrum shows a step-like increase of the event rate at the energy of each electron shell for increasing recoil energies.
In order to include the detector resolution with these steps, the Compton spectrum \cite{Brown:2014gza,Barker:2016eil} is described by a sum of error functions as
\begin{equation} \label{eq:compton}
B_\text{C}(E_\text{ee}) = a_\text{C} \sum \limits_{j = \text{K,L,M}} 0.5 A_j\left(1+\text{erf}\left(\frac{E_\text{ee}-\mu_j}{\sqrt{2}\sigma_j}\right)\right),
\end{equation}
where $a_\text{C}$ is the amplitude of the Compton spectrum and is a free parameter to be estimated in our analysis, $\mu_j$ and $\sigma_j$ are the energy and observed resolution of the Ge K-shell, M-shell, or L-shell activation peaks as listed in Tab.~\ref{tab:activation_peaks}, and $A_j$ are the electron shell-dependent constants given as $A_\text{K}=0.0041\pm0.0001$, $A_\text{L}=0.015\pm0.001$, and $A_\text{M}=0.0345\pm0.0006$ as determined by a fit to calibration data \cite{Barker_thesis}. 

Secondly, the Ge crystal has tritium contamination from above-ground exposure to cosmic rays (prior to installation underground); background events result from the $\beta^-$~decays of these tritium atoms.
The emitted electron has a spectrum given for $E_\text{e}< Q$ by
\begin{eqnarray} \label{eq:tritium}
B_\text{T}(E_\text{e}) &=& a_\text{T} (Q-E_\text{e})^2 (E_\text{e}+m_e) \nonumber \\
& & \times F(Z_\text{He},E_\text{e}) \sqrt{E_\text{e}^2+2E_\text{e} m_e},
\end{eqnarray}
where $a_\text{T}$ is the amplitude of the tritium spectrum and is a free parameter to be estimated by our analysis, \mbox{$Q=18.6$\,keV$_\text{ee}$} is the endpoint energy of the tritium decay, $m_e$ is the electron mass, and $F(Z_\text{He},E_\text{e})$ is the Fermi function.
Because $Q\ll m_e$, we can approximate
\begin{equation}
F(Z_\text{He},E_\text{e})\simeq \frac{2\pi\eta}{1-e^{-2\pi\eta}}
\end{equation}
with
\begin{equation}
\eta=\frac{\alpha Z_\text{He} E_\text{e}}{p_\text{e}}.
\end{equation}
Here, $\alpha$ is the fine structure constant, $Z_\text{He}=2$ is the atomic number of the tritium decay daughter nucleus, and $E_\text{e}$ and $p_\text{e}$ are the kinetic energy and momentum of the emitted electron \cite{Cohen1971}. 
The results presented here will be compared with a dedicated analysis of the tritium and Compton background levels for CDMSlite Run 2 \cite{Agnese:2019app}.

\section{Maximum Likelihood Effective Field Theory Analysis} \label{sec:MLE}

The EFT framework expands the number of interactions considered between dark matter particles ($\chi$) and nucleons ($N$) by including spin-independent (SI), multiple spin-dependent (SD), angular-momentum-dependent (LD), and combinations of angular-momentum-and-spin-dependent (LSD) interactions for a total of 10 interaction types \cite{Anand:2013yka,Fitzpatrick:2012ib,Fitzpatrick:2012ix}.
The effective interaction Lagrangian to describe elastic DM-nucleon scattering can be expressed as 
\begin{equation} \label{eq:EFT_lagrangian}
\mathcal{L}_\text{EFT} = \sum_\tau \sum_i c_i^\tau \mathcal{O}_i \overline{\chi} \chi \overline{\tau} \tau,
\end{equation}
where $\tau$ can either be a sum over proton and neutron interactions or over isoscalar and isovector interactions.
The sum over $i$ is attributed to the sum over all interaction types which are referred to by the operators $\mathcal{O}_i$.
Following the procedure as outlined in Ref.~\cite{Rogers:2016jrx}, the coupling coefficients $c_i$ can be converted between the nucleon (neutron $n$, proton $p$) and isospin bases using
\begin{equation} \label{eq:coefficients}
\begin{split}
&c_i^0 = \frac{1}{2} \left( c_i^p + c_i^n \right) \\
&c_i^1 = \frac{1}{2} \left( c_i^p - c_i^n \right).
\end{split}
\end{equation}
In this analysis the isospin basis is used, where $c_i^0$ is the isoscalar and $c_i^1$ is the isovector interaction \cite{Anand:2013yka}. 

A summary of the EFT operators arranged by parity (P), time-reversal symmetries (T) and spin dependence ($\vec{S}_\chi$) is given in Tab.~\ref{tab:EFT_operators}.
The kinematic variables follow the definitions in Ref.~\cite{Fitzpatrick:2012ix}.
The relative incoming velocity between DM and nucleon, $\vec{v} = \vec{v}_\chi - \vec{v}_\text{N}$, is expressed in a Hermitian way as the transverse component defined as $\vec{v}^\bot \equiv \vec{v} + \frac{\vec{q}}{2\mu_\text{N}}$ with $\vec{q}$ being the momentum transfer and $\mu_\text{N} = \frac{m_\chi m_\text{N}}{m_\chi + m_\text{N}}$ the reduced mass of the DM-nucleon system.
The parameters $\vec{S}_\text{N}$ and $m_\text{N}$ are the nucleon's spin and mass.

\begin{table}[ht!]
\caption{EFT interaction operators separated into categories of similar parity (P), WIMP spin ($\vec{S}_{\chi}$) dependence and behaviour under time reversal transformation (T) \cite{Fitzpatrick:2012ix}.
The interactions are further categorized as spin-independent (SI), spin-dependent (SD), angular-momentum-dependent (LD) or a combination of angular-momentum-dependent and spin-dependent (LSD), where in these four cases the spin dependence is associated with the nucleon target's spin ($\vec{S}_\text{N}$).
The right-most column states the functional form of the EFT operators based on the relevant kinematic parameters.
$\mathcal{O}_2$ is included in the table for historical purposes but is not considered in this analysis because it is not a reduced form of a relativistic operator \cite{Fitzpatrick:2012ib}.}
\label{tab:EFT_operators}
\begin{ruledtabular}
\begin{tabular}{lcc}
\multicolumn{3}{c}{P-even, $\vec{S}_{\chi}$-independent, T-conserving} \\
$\mathcal{O}_1$ & Standard SI & $1$ \\
$\mathcal{O}_2$ & Not considered & $(\vec{v}^{\bot})^2$ \\
$\mathcal{O}_3$ & LSD and SD & $i \vec{S}_\text{N} \cdot ({\vec{q} / m_\text{N}} \times \vec{v}^{\bot})$ \\
\colrule
\multicolumn{3}{c}{P-even, $\vec{S}_{\chi}$-dependent, T-conserving} \\
$\mathcal{O}_4$ & Standard SD & $\vec{S}_{\chi} \cdot \vec{S}_\text{N}$ \\
$\mathcal{O}_5$ & SI and LD & $i \vec{S}_{\chi} \cdot (\vec{q}/m_\text{N} \times \vec{v}^{\bot})$ \\
$\mathcal{O}_6$ & SD & $(\vec{S}_{\chi} \cdot \vec{q}/m_\text{N})(\vec{S}_\text{N} \cdot \vec{q} / m_\text{N})$ \\
\colrule
\multicolumn{3}{c}{P-odd, $\vec{S}_{\chi}$-independent,T-conserving} \\
$\mathcal{O}_7$ & SD & $\vec{S}_\text{N} \cdot \vec{v}^{\bot}$ \\
\colrule
\multicolumn{3}{c}{P-odd, $\vec{S}_{\chi}$-dependent, T-conserving} \\
$ \mathcal{O}_8$ & SI and LD & $\vec{S}_{\chi} \cdot \vec{v}^{\bot}$ \\
$ \mathcal{O}_9$ & SD & $i \vec{S}_{\chi} \cdot (\vec{S}_\text{N} \times \vec{q} /m_\text{N})$ \\
\colrule
\multicolumn{3}{c}{P-odd, $\vec{S}_{\chi}$-independent, T-violating} \\
$\mathcal{O}_{10}$ & SD & $i \vec{S}_\text{N} \cdot \vec{q} / m_\text{N} $ \\
\colrule
\multicolumn{3}{c}{P-odd, $\vec{S}_{\chi}$-dependent, T-violating} \\
$\mathcal{O}_{11}$ & SI & $i \vec{S}_{\chi} \cdot \vec{q} / m_\text{N} $ \\
\end{tabular}
\end{ruledtabular}
\end{table}

From the operators listed in Tab.~\ref{tab:EFT_operators}, the traditional SI and SD operators are $\mathcal{O}_1$ and $\mathcal{O}_4$, respectively.
Each EFT operator has both isoscalar and isovector (or proton/neutron) components and is found as a leading-order term in the non-relativistic reduction of a relativistic operator with a traditional spin-0 or spin-1 mediator, except for $\mathcal{O}_{2}$.
For this reason, we do not consider $\mathcal{O}_{2}$ in our analysis. 
Interference exists between isoscalar and isovector terms within a single operator and can exist between operators within the same symmetry categories.
In particular there can be interference between operators $\mathcal{O}_{1}$ and $\mathcal{O}_{3}$, among $\mathcal{O}_{4}$, $\mathcal{O}_{5}$, and $\mathcal{O}_{6}$, and between $\mathcal{O}_{8}$ and $\mathcal{O}_{9}$~\cite{Fitzpatrick:2012ix}.
The experimental response to DM-nucleon scattering is sensitive to the specific EFT operator. 
Momentum-independent operators ($\mathcal{O}_{1}$, $\mathcal{O}_{4}$, $\mathcal{O}_{7}$, and $\mathcal{O}_{8}$) lead to an exponential energy spectrum, where most of the events are expected in the lowest-energy bins.
In contrast, operators with an explicit momentum dependence ($\mathcal{O}_{3}$, $\mathcal{O}_{5}$, $\mathcal{O}_{6}$, $\mathcal{O}_{9}$, $\mathcal{O}_{10}$, and $\mathcal{O}_{11}$) produce a spectrum which resembles that of $\beta$-decays.
In this case the spectrum's characteristics are a vanishing rate at zero recoil energy and a dependence of the maximum rate on the DM particle's mass.

In order to interpret direct DM detection data within the EFT framework, a Poissonian likelihood is defined and binned over energy.
For $n$ energy bins, the DM-only likelihood, $\mathcal{L}_\text{DM}$, is given by
\begin{equation}\label{eq:Poisson}
\mathcal{L}_\text{DM}(\{m_\chi,c^0_i,c^1_i\} \vert \vec N) = \prod_{k=1}^n \frac{1}{N_k!} \eta_k^{N_k} e^{-\eta_k},
\end{equation}
where $N_k$ is the observed number of events in each bin.
The parameter $\eta_k$ is the number of events theoretically expected in the $k^{\rm th}$ energy bin based on the efficiency (set to zero in excluded ranges), the live time of the experiment and the EFT model(s) chosen for the analysis. 

For example, in a fit to a single EFT operator model, the parameters of Eq.~(\ref{eq:Poisson}) consist of $\{m_\chi, c_i^0, c_i^1\}$: the DM particle's mass ($m_\chi$), along with the isoscalar ($c_i^0$) and isovector ($c_i^1$) coupling coefficients of the chosen EFT operator $\mathcal{O}_{i}$.
Then the theoretical number of events in any given bin is calculated as an integral over $E_k$, the recoil energy of each bin, by
\begin{equation}
\eta_k(\{m_\chi, c_i^0,c_i^1\})= \int_{E_k} \epsilon(E_\text{R}) \frac{dR} {dE_\text{R}} dE_\text{R},
\end{equation}
where $\epsilon(E_\text{R})$ is the energy-dependent efficiency times the experiment's exposure.
The differential event rate, ${dR}/{dE_\text{R}}$, also depends on $m_\chi$ and the coupling coefficients.
If the energy dependence of the expected background is known, it can be included in the likelihood, leading to the combined likelihood of
\begin{equation}
\mathcal{L}_\text{DM+bkg}(\{m_\chi,c^0_i,c^1_i\} \vert \vec{N}) = \prod_{k=1}^n \frac{1}{N_k!} (\eta_k + b_k)^{N_k} e^{-(\eta_k+b_k)},
\end{equation}
where $b_k$ is the expected number of background events in the $k^{\rm th}$ energy bin.
The value of $b_k$ would be determined using the amplitudes of the relevant background parameters (i.e. Compton and/or tritium backgrounds) and $\epsilon(E_\text{R})$ \cite{Rogers:2016jrx}.

In order to determine which model or EFT operator gives the best fit to the data set, the Bayesian evidence can be calculated for each likelihood.
The Bayesian evidence, $\mathcal{Z}$, appears in Bayes' equation,
\begin{equation}
\mathcal{P}(\{m_\chi,c^0_i,c^1_i\})=\frac{\mathcal{L}(\{m_\chi,c^0_i,c^1_i\} \vert \vec{N}) \text{Pr}(\{m_\chi,c^0_i,c^1_i\})}{\mathcal{Z}_i},
\end{equation}
as the normalization constant, where $\mathcal{P}(\{m_\chi,c^0_i,c^1_i\})$ is the posterior distribution and $\text{Pr}(\{m_\chi,c^0_i,c^1_i\})$ is the prior \cite{Feroz:2008xx,Feroz:2007kg}.
Therefore, the evidence for a single EFT operator, $\mathcal{O}_i$, is given by
\begin{equation} \label{eq:evidence}
\mathcal{Z}_i=\int dm_\chi dc^0_i dc^1_i \mathcal{L}(\{m_\chi,c^0_i,c^1_i\} \vert \vec{N}) \text{Pr}(\{m_\chi,c^0_i,c^1_i\}).
\end{equation}
For a given set of models, if all are considered equally probable \textit{a priori}, the ratio between overall posterior probabilities of any two models is equal to the ratio between their two Bayesian evidences.
Thus ratios of Bayesian evidences indicate the relative plausibility of each model in light of the data.

An efficient method of scanning over all possible dimensions, especially considering the likelihood functions tend to be multimodal, is by using the nested sampling Monte Carlo software package \textsc{MultiNest} \cite{Feroz:2013hea} -- a Bayesian inference tool for parameter estimation as well as model comparison and selection.
The nested sampling technique used by \textsc{MultiNest} involves an optimized set of sample points from the full likelihood.
At each iteration of the algorithm the sample point of the lowest likelihood is replaced with a point of higher likelihood which yields the points of highest likelihood in an iterative way.

In our analysis the single operator models result in 3D to 5D likelihoods depending on the number of background parameters considered.
Assuming that the background amplitudes are independent from the parameters describing the EFT operators (WIMP mass and coupling coefficients), this likelihood can be split into one 3D likelihood describing the EFT operator and one likelihood describing the background.
By integrating over one parameter of the EFT likelihoods, one obtains 2D marginalized likelihoods which are used to obtain 95\% credibility contours for each plane of the 2D marginalization in Sec.~\ref{sec:likelihoods}.
As a final step, artifacts caused by poor sampling in the 2D marginalized likelihoods with flat regions are smoothed using a Savitzky-Golay filter.

Instead of calculating the EFT likelihoods in terms of the coupling coefficients, we made a change of variables to coupling-coefficient amplitude, $A_i$, and coupling-coefficient phase, $\theta_i$.
These polar coordinates are a natural choice for the EFT parameter space because the interaction cross section, $\sigma_i$, is approximately proportional to the square of the coupling-coefficient amplitude, $\sigma_i \propto A_i^2$.
Because there is also no shape dependence of the differential rate on $A_i$, only the WIMP mass and $\theta_i$ determine the shape of the differential rate spectrum.
The conversion from polar coordinates into the coupling-coefficient plane is achieved with
 \begin{equation}
 \begin{split}
 &c^0_i=A_i \sin(\theta_i), \\
 &c^1_i = A_i \cos(\theta_i).
 \end{split}
 \end{equation}

The coupling-coefficient amplitude is included with the prior probability function
\begin{eqnarray} \label{eq:flat_logA}
\text{Pr}(\log A_i) &=& \left\{ \begin{array}{l}
    {\frac{1}{\log(A_{i,\text{max}})-\log(A_{i,\text{min}})}} \\
    {0,\text{ otherwise}} \\
    \end{array} \right. \\
    & & \text{with } ~A_{i,\text{min}} \leq A_i \leq A_{i,\text{max}}, \nonumber
 \end{eqnarray}
where $A_{i,\text{max}}$ and $A_{i,\text{min}}$ were chosen for each operator to span roughly 13 orders of magnitude around a lower number of expected events that still allows for the observed spectrum to be a WIMP signal. 

The coupling-coefficient phase is included with a flat prior given by
 \begin{equation}
 \text{Pr}(\theta_i) = \left\{ \begin{array}{l}
    {\frac{1}{\pi}, ~-\pi/2 \leq \theta_i \leq \pi/2} \\
    {0,\text{ otherwise}} \\
    \end{array} \right. .
\end{equation}
The limits of $(-\pi/2 \leq \theta_i \leq \pi/2)$ were chosen because only the relative sign of the coupling coefficients $c_i^0$ and $c_i^1$ matter.
Both positive and both negative (or either one being positive and the other negative) have the same physical meaning, hence, all possible physical interactions are included within the chosen prior.
In the equivalent formalism using the nucleon basis (cf.~Eq.~(\ref{eq:coefficients})) the prior choice for $\theta_i$ would be of uniform character as well.

The WIMP mass is included using a flat prior ranging from 0 to 25\,GeV/$c^2$.
As CDMSlite Run 2 was designed to focus on the detection of low-mass WIMPs, the prior for $m_\chi$ was chosen with an upper limit of $25$\,GeV/$c^2$ for all operators.
This mass limit was chosen because for all operators with a $\beta$-decay-like spectral shape above $25$\,GeV/$c^2$, a large portion of the theoretical event rate spectrum is past the 60\,keV$_\text{nr}$ upper threshold.
Additionally, data from previous CDMS Ge experiments have set strong limits in this region on operators with an exponential spectral shape \cite{Agnese:2014aze,Agnese:2015ywx}.
 
For the WIMP mass and coupling-coefficient amplitude, two possible options for the prior probability function are flat or log-flat.
Changing the priors on $m_\chi$ and $A_i$ between these two options can alter the 95\% credibility contours in the amplitude-mass plane by a factor of 1.0 to $\sim 2.5$.
In comparison to the 13 orders of magnitude scanned for $A_i$, this change is fairly minor. 
Therefore, the prior probability distributions stated above were chosen because of their ability to sufficiently sample the ROI for each parameter.
Further discussion on the effect of the choice of prior probability distribution can be found in Appendix \ref{app:priors}.
 
Each background amplitude was included in the likelihood analysis using a flat prior spanning from zero to the value corresponding to a background rate of two times the total number of detected events.
In this way, every calculation that includes backgrounds allows for the possibilities of the detected events containing no background or only background events.

\section{Comparison to Published CDMSlite Run 2 Limit} \label{sec:compare}

\begin{figure*}[ht!]
   \centering
   \includegraphics[width=\columnwidth]{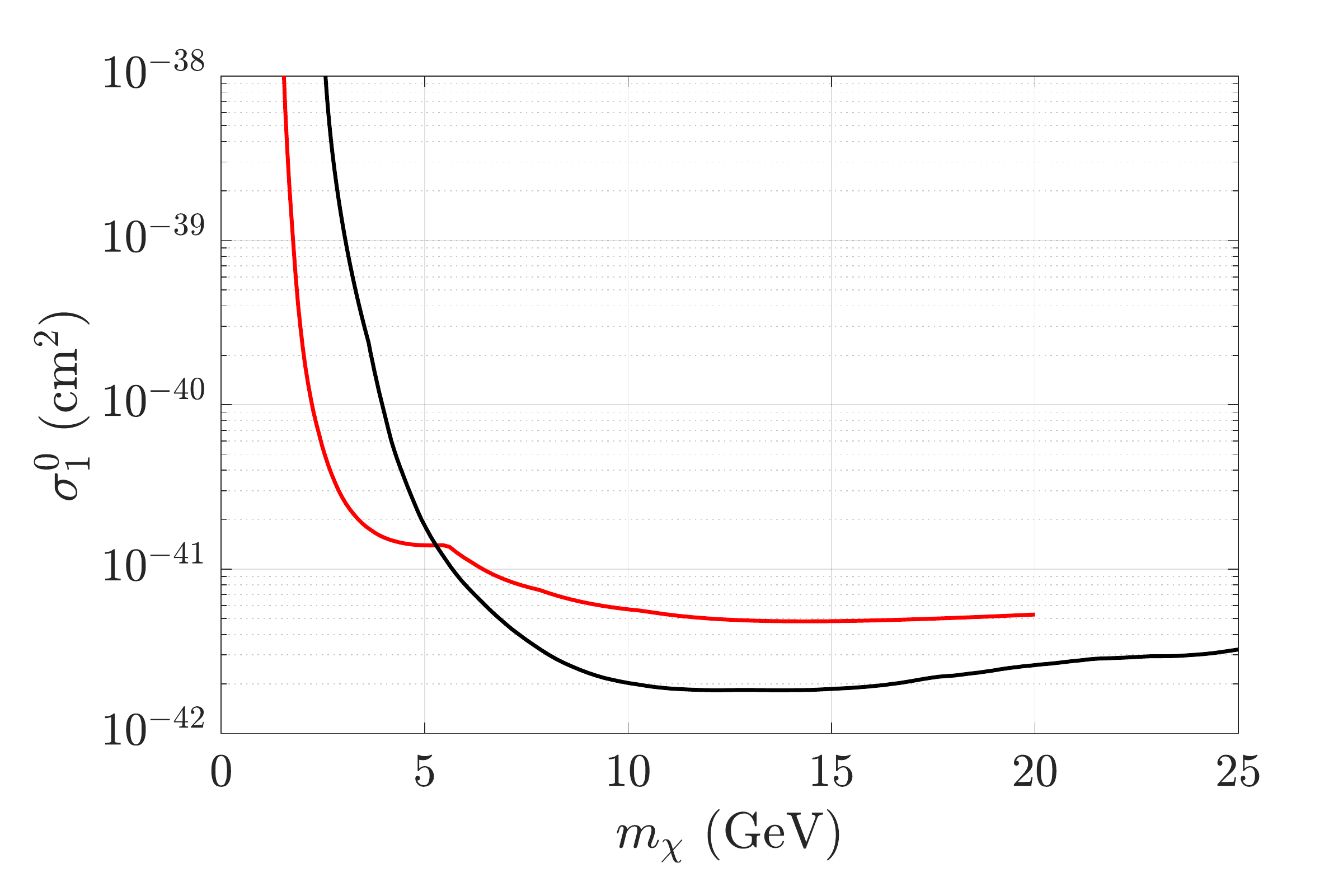}
   \includegraphics[width=\columnwidth]{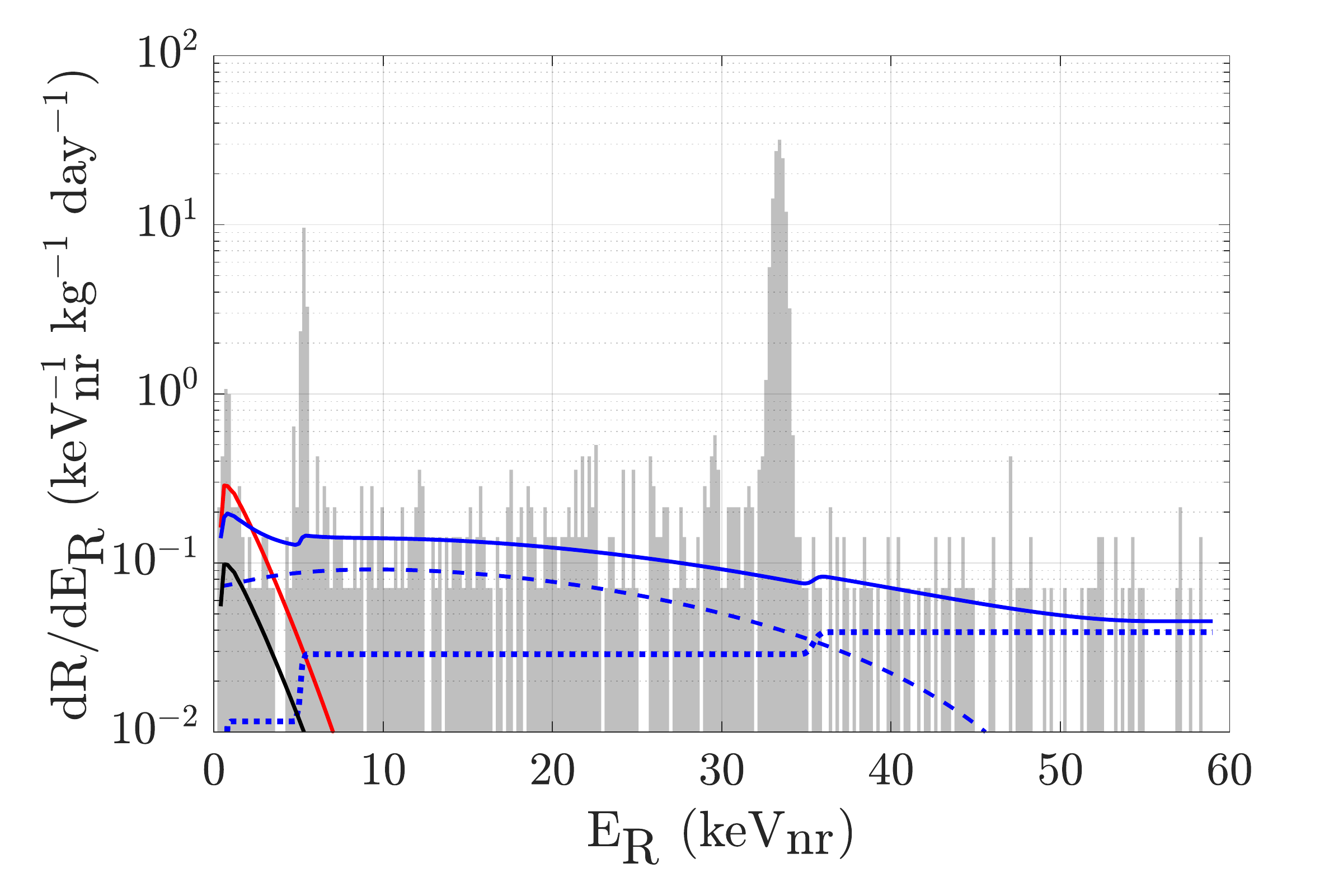}
   \caption{Left: Comparison of the published CDMSlite Run 2 Optimum Interval limit at 90\% C.L. \cite{Agnese:2015nto} (red solid line) and the EFT isoscalar operator $\mathcal{O}_1$ upper limit at 95\% credibility (black solid line). Right: Comparison of the observed differential recoil energy spectrum to the signal and background models. The solid red (solid black) line shows the upper limit on the expected signal contribution from a 10\,GeV/c$^2$ DM particle using the OI method (the EFT Bayesian upper limit for $\mathcal{O}_1$). The blue dotted (dashed) line represents the Compton (tritium) background and the blue solid line is the sum of the EFT signal and total background.}
   \label{fig:OI_comparison}
\end{figure*}

As a test of the methodology, and to show the power of including a background model in the likelihood, a 2D posterior distribution was determined by marginalising over the background parameters for EFT operator $\mathcal{O}_1$ using the WIMP mass and the isoscalar coupling-coefficient amplitude ($\theta = 0$).
This posterior distribution was calculated using \textsc{MultiNest} and the background models excluding the energy regions of the known activation peaks as described above.
The overall analysis procedure and marginalization of the multi-dimensional likelihoods, which result in contours representing the 95\% Bayesian credible regions in 2D planes of $m_\chi$ and $A_i$, are described in Ref.~\cite{Rogers:2016jrx}.
Similarly, a 1D marginalized likelihood can be used to determine the 95\% credibility regions for each parameter individually, by integrating down from the point of highest likelihood.

The resulting 95\% credibility contour can be compared to the published CDMSlite Run~2 Optimum Interval (OI) limit on the spin-independent WIMP-nucleon elastic scattering cross section \cite{Agnese:2015nto} as shown in Fig.~\ref{fig:OI_comparison}.
This direct comparison is possible, because for operator $\mathcal{O}_1$, cross section ($\sigma^0_1$) and coupling-coefficient amplitude ($A_1$) can be converted as 
\begin{equation}
\sigma_1^0 = \frac{A^2 m_\text{N}^2}{4\pi\langle V \rangle^4(1+A)^2}A_1,
\end{equation}
where $A$ is the number of nucleons of the target material, $m_\text{N}$ is the effective nucleon mass and $\langle V \rangle = 246.2$\,GeV/$c^2$ is the Higgs vacuum expectation value, used here to represent the electroweak scale and to define dimensionless coefficients \cite{Anand:2013yka}.

The CDMSlite Run 2 OI limit calculation and the EFT contour calculation are quite different.
The EFT contour is the result of a Bayesian analysis, while the OI calculation is a more traditional method to derive a limit at a chosen confidence interval \cite{Yellin:2002xd}.
The Bayesian EFT limit is calculated at 95\% credibility whereas the OI limit has been reported at a confidence level of 90\%.
Furthermore, the OI limit calculation does not include any background models; i.e., it is assumed that the measured spectrum could be due entirely to WIMPs.
For the EFT contours, the Compton scattering and tritium contamination spectra (cf.~Eq.~(\ref{eq:compton}) and (\ref{eq:tritium})) were included separately as $a_\text{C}$ and $a_\text{T}$ in the likelihood analysis.
The maximum recoil energies considered vary between the two analyses as well.
The analysis stopped at 20\,keV$_\text{nr}$ for the OI method and at 60\,keV$_\text{nr}$ for the EFT analysis.

Furthermore, the EFT limit does not reach as low in terms of excluded WIMP mass range (\mbox{$\sim 1.3$\,GeV/$c^2$} for the OI analysis and $\sim 2.6$\,GeV/$c^2$ for the EFT method).
This is due to the raised energy threshold from removing energy regions around the known activation peaks.
At higher recoil energies, the EFT likelihood analysis is better able to take into account the different shapes of the data and input spectra.
As a result, the EFT limit reaches a cross section about three times lower at 10\,GeV/$c^2$ compared to the OI result.
For WIMPs heavier than 5\,GeV/$c^2$ the inclusion of a background model and the higher maximum recoil energy considered result in stronger exclusion limits.
The stronger EFT exclusion limits also hold qualitatively when considering the differently chosen level of statistical credibility (cf.~Fig.~\ref{fig:OI_comparison}).

For completeness, a comparison of the WIMP spectra corresponding to the cross sections at 10\,GeV/$c^2$ is shown in Fig.~\ref{fig:OI_comparison}.
For the EFT analysis, the included background models are shown as well.
Both the OI and the EFT spectra are compared against the same CDMSlite Run~2 data.

\section{Bayesian Evidence and Model Selection} \label{sec:evidence}

As defined in Eq.~(\ref{eq:evidence}), the Bayesian evidence can be used to select which model is most likely to describe the observed DM search data.
The higher the evidence, the better is the fit of the corresponding model to the data.
Each combination of EFT operator and background configuration can be considered a separate model.
The models considered are:
\begin{itemize}
\item \textit{\textbf{Two-parameter}} background model -- background model includes Compton and tritium backgrounds separately with two parameters ($a_\text{C}$ and $a_\text{T}$) for a total of five parameters ($m_\chi$, $A_i$, $\theta_i$, $a_\text{C}$, and $a_\text{T}$).
\item \textit{\textbf{Compton only}} -- background model includes Compton background only ($a_\text{C}$) for a total of four parameters ($m_\chi$, $A_i$, $\theta_i$, and $a_\text{C}$).
\item \textit{\textbf{Tritium only}} -- background model includes tritium backgrounds only ($a_\text{T}$) for a total of four parameters ($m_\chi$, $A_i$, $\theta_i$, and $a_\text{T}$).
\item \textit{\textbf{No background}} -- no background models included for a total of three parameters ($m_\chi$, $A_i$, and $\theta_i$).
\item Two-parameter \textit{\textbf{background-only}} -- described with two background parameters ($a_\text{C}$ and $a_\text{T}$) but no signal parameters.
\end{itemize}

\subsection{Comparison Between Models}

\begin{table*}[ht!]
\caption{Ratio of Bayesian evidence of each model compared to the reference, $\mathcal{Z}_\text{model}/\mathcal{Z}_\text{ref}$. The reference chosen is the two-parameter background-only model (boldface).}
\label{tab:evidence_ratio}
\begin{ruledtabular}
\begin{tabular}{ccccc}
 Bkg Model & \textit{\textbf{two-parameter}} & \textit{\textbf{Compton only}} & \textit{\textbf{tritium only}} & \textit{\textbf{no background}} \\
 Dimensions & 5 (Bkg = 2) & 4 (Bkg= 1) & 4 (Bkg = 1) & 3 (Bkg = 0) \\
 \colrule
 EFT Models & \multicolumn{4}{c}{$\mathcal{Z}_\text{model}/\mathcal{Z}_\text{ref}$} \\
 \colrule
 Bkg only & $\displaystyle \bf{1.00}$ & $(4.2\pm0.2)\times10^{-13}$ & $(2.05\pm0.08)\times10^{-100}$ & -- \\
 $\mathcal{O}_{1}$ & $0.69\pm0.03$ & $(6.9\pm0.4)\times10^{-6}$ & $(4.4\pm0.3)\times10^{-18}$ & $(6.5\pm0.4)\times10^{-16}$ \\
 $\mathcal{O}_{3}$ & $0.40\pm0.02$ & $(3.9\pm0.2)\times10^{-4}$ & $0.027\pm0.001$ & $(2.2\pm0.1)\times10^{-11}$ \\
$\mathcal{O}_{4}$ & $0.55\pm0.02$ & $(1.74\pm0.09)\times10^{-6}$ & $(7.1\pm0.5)\times10^{-8}$ & $(5.1\pm0.3)\times10^{-5}$ \\
$\mathcal{O}_{5}$ & $0.45\pm0.02$ & $0.0138\pm0.0007$ & $0.0103\pm0.0005$ & $0.0089\pm0.0005$ \\
$\mathcal{O}_{6}$ & $0.37\pm0.02$ & $(3.7\pm0.2)\times10^{-4}$ & $0.023\pm0.001$ & $(1.12\pm0.07)\times10^{-9}$ \\
$\mathcal{O}_{7}$ & $0.59\pm0.03$ & $(2.5\pm0.1)\times10^{-7}$ & $(8.0\pm0.6)\times10^{-22}$ & $(5.5\pm0.3)\times10^{-5}$ \\
$\mathcal{O}_{8}$ & $0.53\pm0.02$ & $(3.2\pm0.2)\times10^{-3}$ & $(3.0\pm0.2)\times10^{-4}$ & $0.0068\pm0.0003$ \\
$\mathcal{O}_{9}$ & $0.54\pm0.02$ & $0.0164\pm0.0008$ & $(9.9\pm0.7)\times10^{-6}$ & $(8.4\pm0.05)\times10^{-5}$ \\
$\mathcal{O}_{10}$ & $0.35\pm0.02$ & $0.0164\pm0.0008$ & $0.0029\pm0.0001$ & $0.0173\pm0.0009$ \\
$\mathcal{O}_{11}$ & $0.47\pm0.02$ & $0.0176\pm0.0009$ & $(5.2\pm0.3)\times10^{-4}$ & $0.019\pm0.001$ \\
\end{tabular}
\end{ruledtabular}
\end{table*}

The relative Bayesian evidence was calculated for each model.
The prior probability functions used to calculate these evidences were a flat mass prior and flat $\log(A_i)$ prior (cf.~Eq.~(\ref{eq:flat_logA})).
It is found that the model with the highest evidence is the background-only model with two parameters; this indicates that the CDMSlite Run~2 data do not have a preference for any non-zero EFT contribution.
Because the two-parameter background-only model has the highest evidence, it was chosen as the reference model for calculating the evidence ratios summarized in Tab.~\ref{tab:evidence_ratio}.
In order to give scale to these ratios, $\log(\mathcal{Z})=-303.40\pm0.03$ for the reference model.

Comparing the \textit{\textbf{two-parameter}} models to the \textit{\textbf{tritium}}, \textit{\textbf{Compton}}, and \textit{\textbf{no background}} models shows that the Bayesian evidence is consistently higher when both background sources are included in the analysis.
This demonstrates that each of the background sources is necessary to describe the CDMSlite data and that no single EFT operator entirely mimics either background shape.
The difference between the predicted signal shape and the background provides sensitivity in a likelihood approach even in a background-limited scenario like for the present CDMSlite Run 2 data set.
In particular, even though the Bayesian evidences show the data are consistent with a background-only model, upper limits on the WIMP mass and the coupling-coefficient amplitudes can be set from the likelihood calculation.

\section{Calculated Posterior Distributions} \label{sec:likelihoods}

The ratios of Bayesian evidences in Tab.~\ref{tab:evidence_ratio} provide an important interim result of our analysis.
However, the main results of interest are the corresponding posterior distributions which will be discussed in the following section. 
In Sec.~\ref{sub:bkg_only} the posterior distribution for the two-parameter background-only model is presented.
We discuss the posterior distributions for the single EFT operator models in Sec.~\ref{sub:single_op}.
It should be noted that also two-operator interference models exist, but they have not been included in this analysis.

\subsection{Background-only Model} \label{sub:bkg_only}

\begin{figure}[ht!]
   \centering
   \includegraphics[width=\columnwidth]{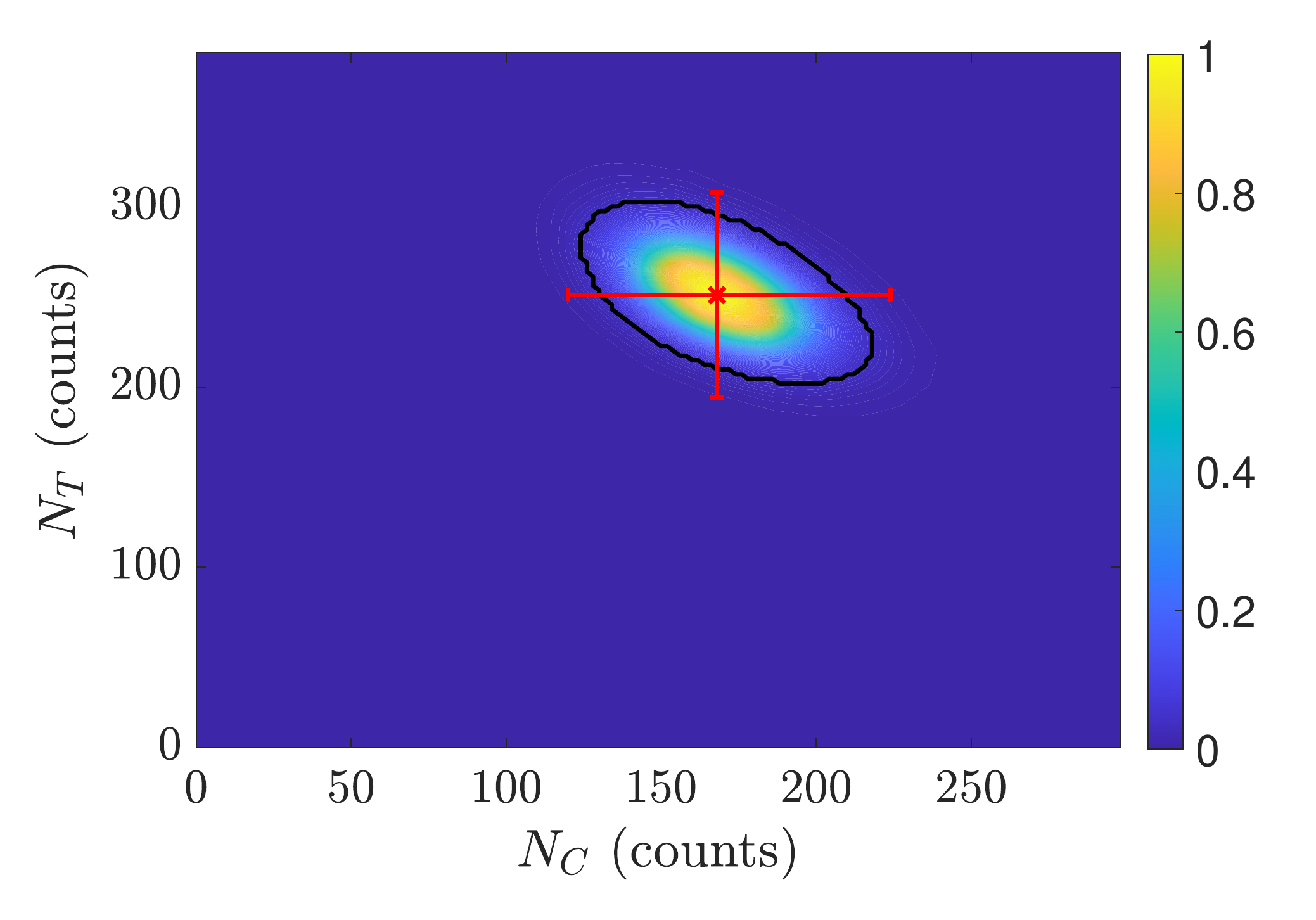}
   \caption{2D posterior distribution for the two-parameter background-only model. The 95\% credibility contour is shown as black line and the best best fit background configuration is shown as a red cross with uncertainties corresponding to 2$\sigma$.}
   \label{fig:twoparameter_bkg_only}
\end{figure}

The total number of background events in the CDMSlite Run 2 data set were estimated using the two-parameter background-only model.
Fig.~\ref{fig:twoparameter_bkg_only} shows the 2D posterior distribution using the total Compton ($N_\text{C}$) and tritium events ($N_\text{T}$) separately.
The total number of events is the sum across all energy bins for the chosen background while including the regions that had been previously removed from the analysis.
We include these regions in order to make direct comparisons between unbinned analyses possible. 
All uncertainties quoted in the following were calculated at the 2$\sigma$ level.

According to the best fit point shown in Fig.~\ref{fig:twoparameter_bkg_only}, the total contribution of all background sources is $420\pm^{80}_{75}$ events.
The total background is composed of \mbox{$N_\text{C} = 168\pm^{56}_{48}$} events due to Compton background and \mbox{$N_\text{T} = 252\pm57$} events attributed to the $\beta^-$ decay of tritium.
The resulting ratio of Compton to tritium events is \mbox{$N_\text{C}/N_\text{T} = 0.67 \pm^{0.27}_{0.24}$}.
As a measure of the systematic uncertainties of the background estimates, a change in the shape of the Compton spectrum (varying parameters in Eq.~(\ref{eq:compton})) on the order of 3\% alters the count of the individual backgrounds on the order of 6\% and the total background on the order of 0.3\%.

In addition, an attempt to calculate the background rates for all background types in CDMSlite Run 2 was completed using a different maximum likelihood technique for the energy range up to $\sim 50$\,keV$_\text{nr}$ \cite{Agnese:2019app}.
This analysis resulted in a calculated total Compton background of $N_\text{C} = 131\pm^{44}_{36}$ events and a total tritium background of $N_\text{T} = 270\pm48$ events leading to a ratio of \mbox{$N_\text{C}/N_\text{T} = 0.48 \pm^{0.18}_{0.16}$}.
By explicitly taking into account the activation peaks in the likelihood fit, the uncertainties on the total background counts turn out to be $\sim 20\%$ smaller compared to the results presented here.
While these two analyses result in slightly different best-fit ratios between Compton to tritium, they both give consistent results for the numbers of events for each species.

\subsection{Single Operator EFT Models} \label{sub:single_op}

\begin{figure*}[ht!]
   \centering
   \includegraphics[width = 0.6\textwidth]{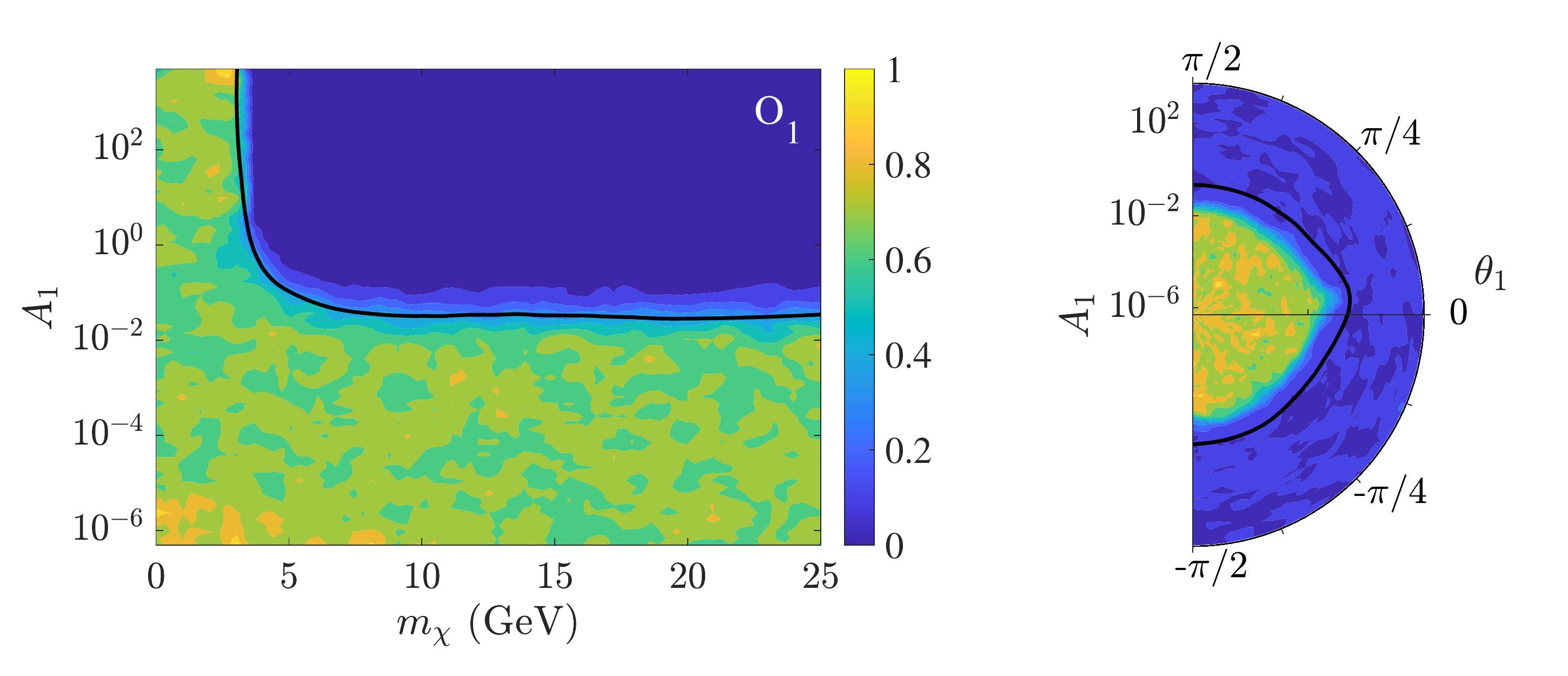}
   \includegraphics[width = 0.6\textwidth]{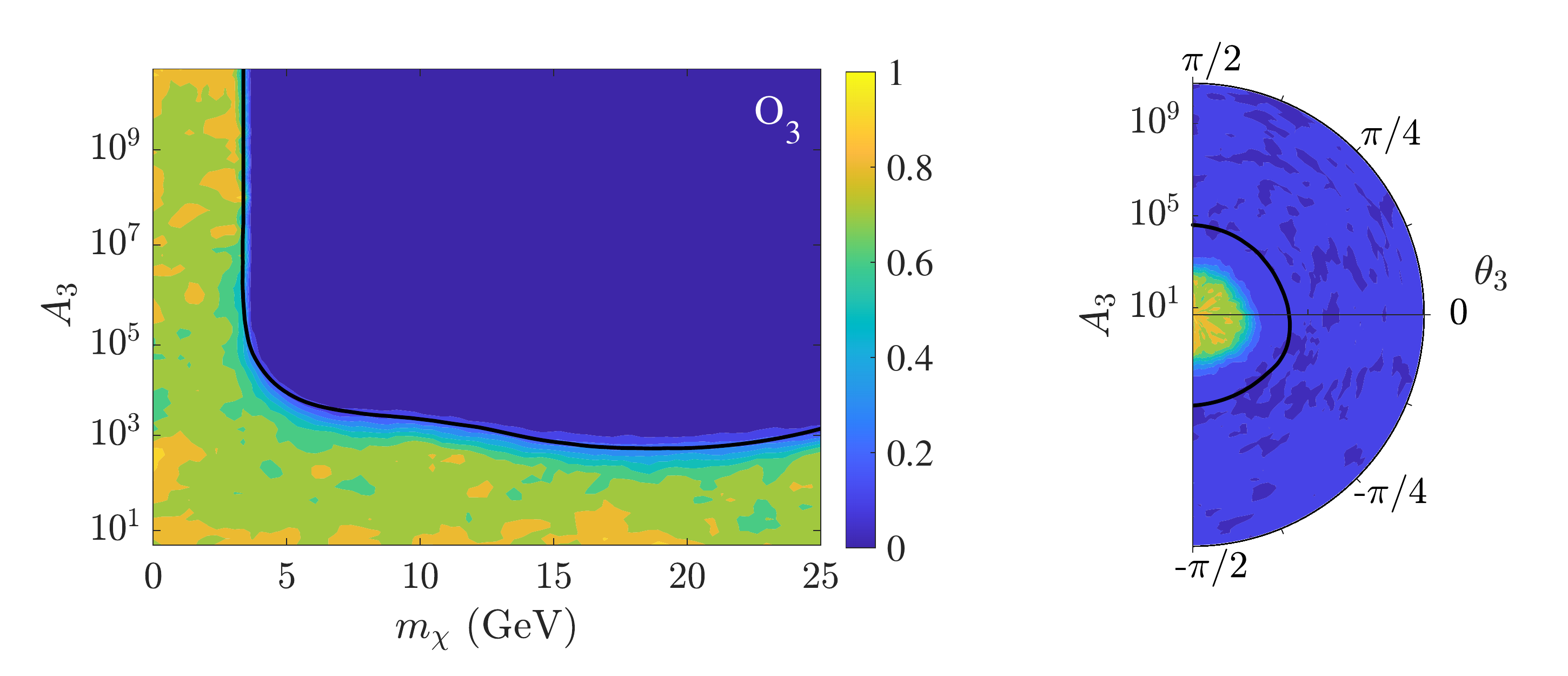}
   \includegraphics[width = 0.6\textwidth]{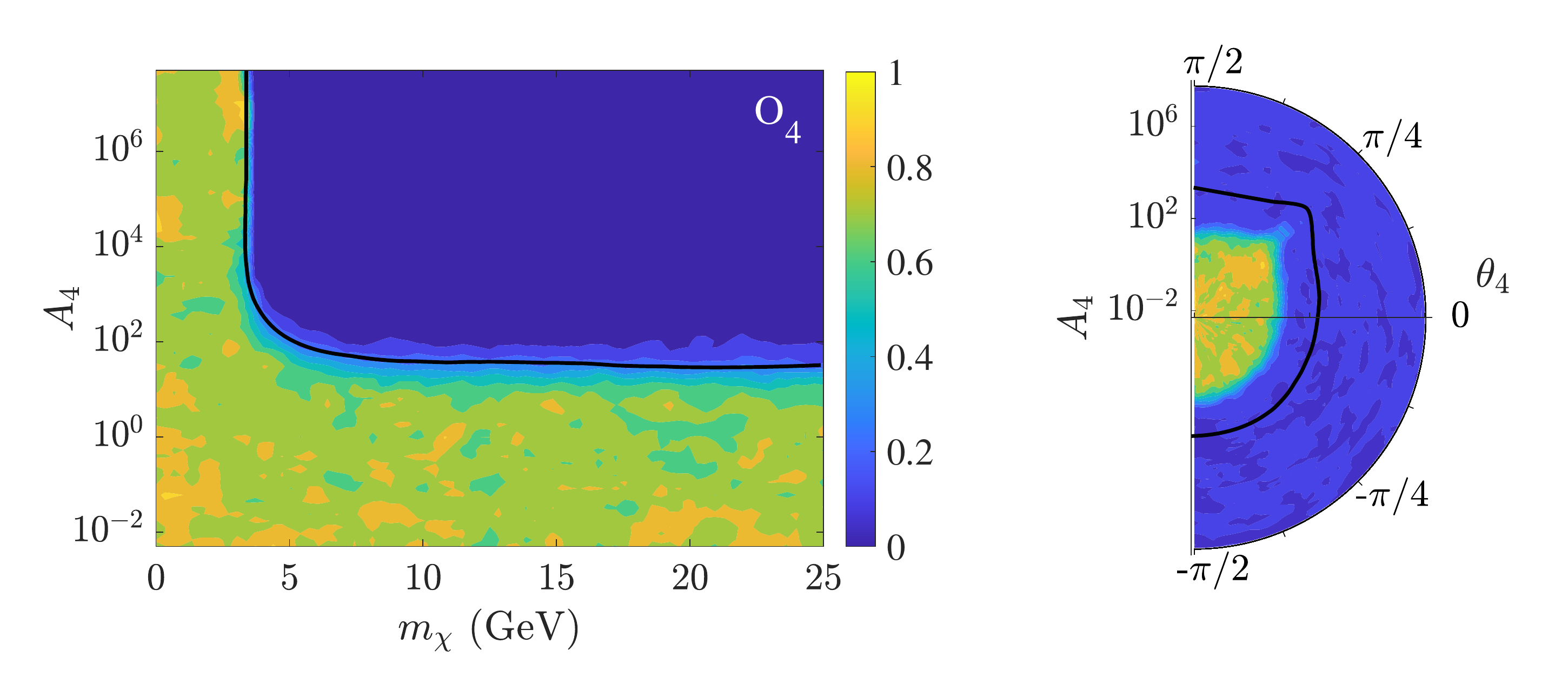}
   \includegraphics[width = 0.6\textwidth]{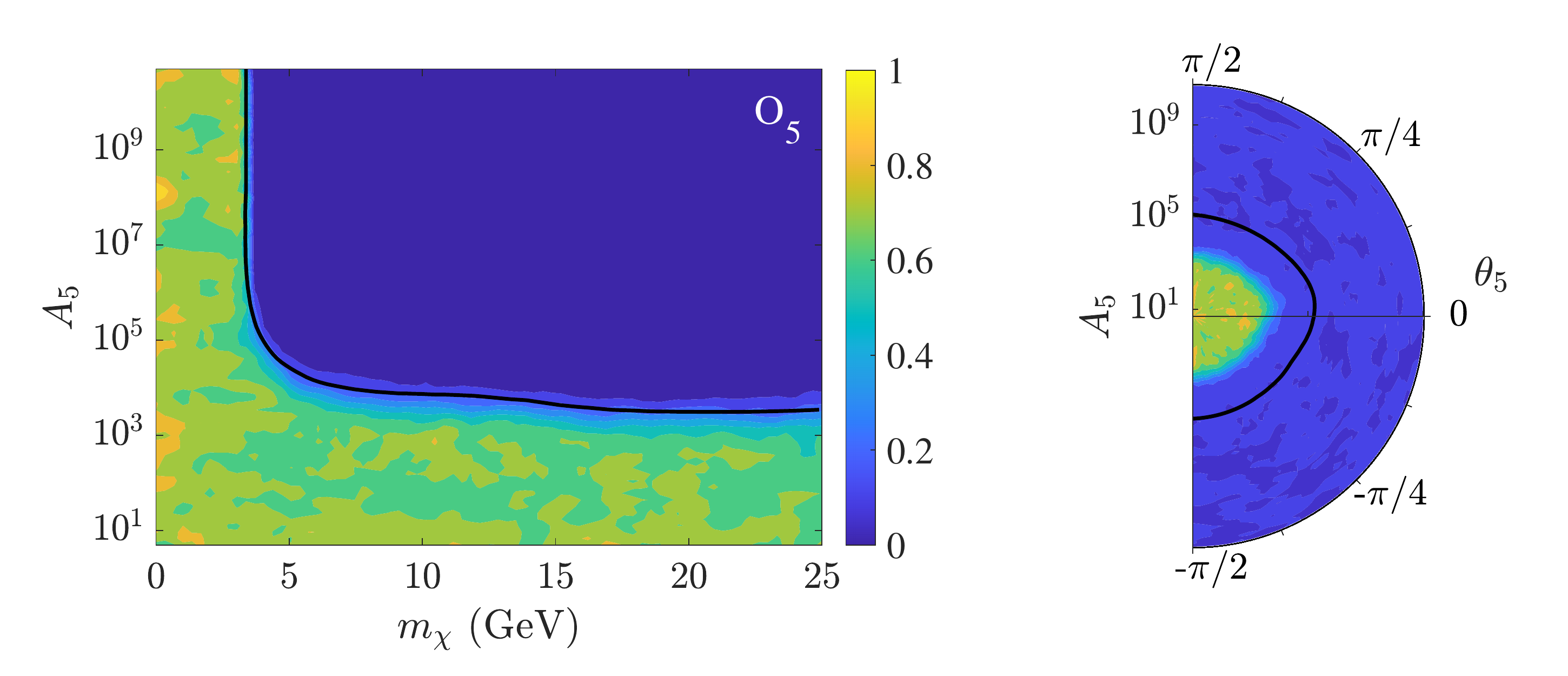}
   \caption{2D marginalized posterior distributions for the polar parameters ($m_\chi$, $A_i$, and $\theta_i$) of each EFT operator ($\mathcal{O}_1$ - $\mathcal{O}_5$) under the two-parameter background model. The respective 95\% credibility contours are shown in black.}
   \label{fig:2D_likelihoods_A}
\end{figure*}

\begin{figure*}[ht!]
   \centering
   \includegraphics[width = 0.6\textwidth]{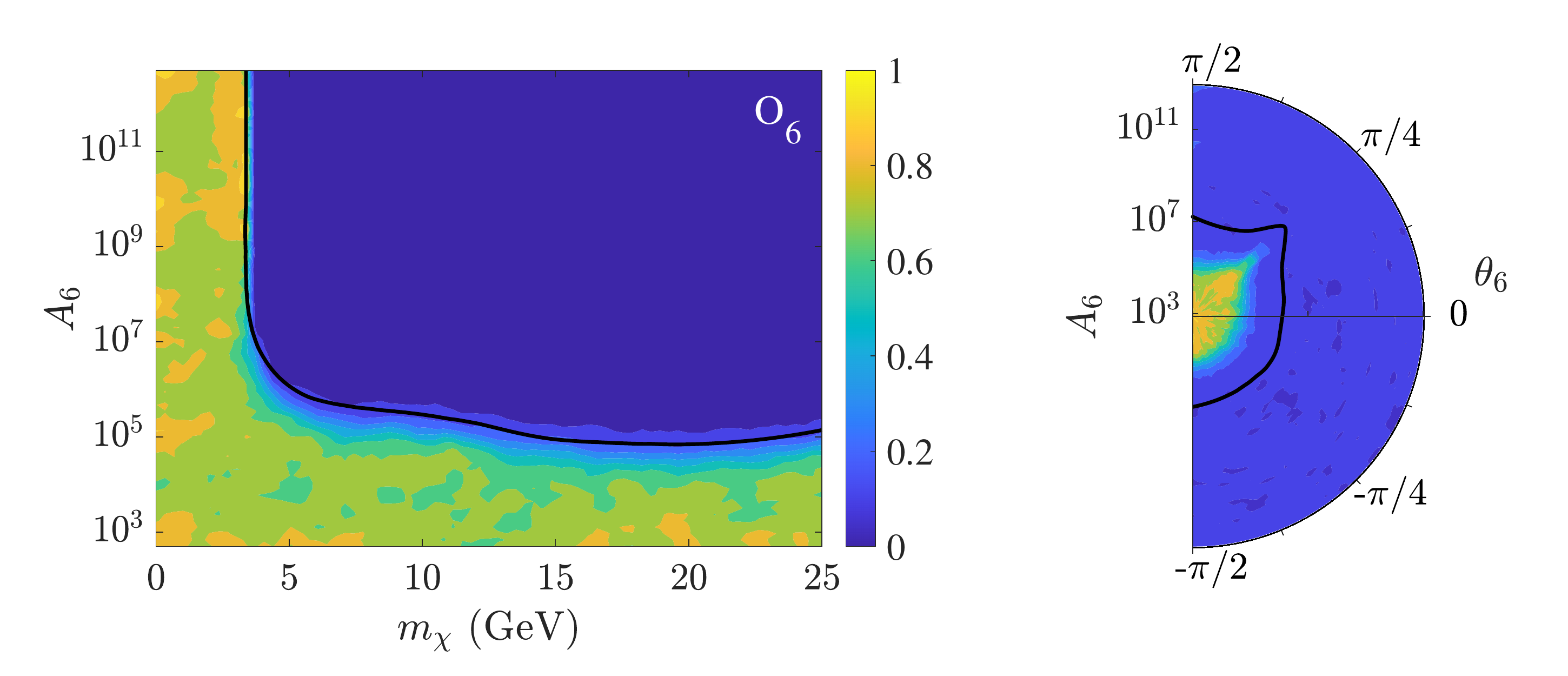}
   \includegraphics[width = 0.6\textwidth]{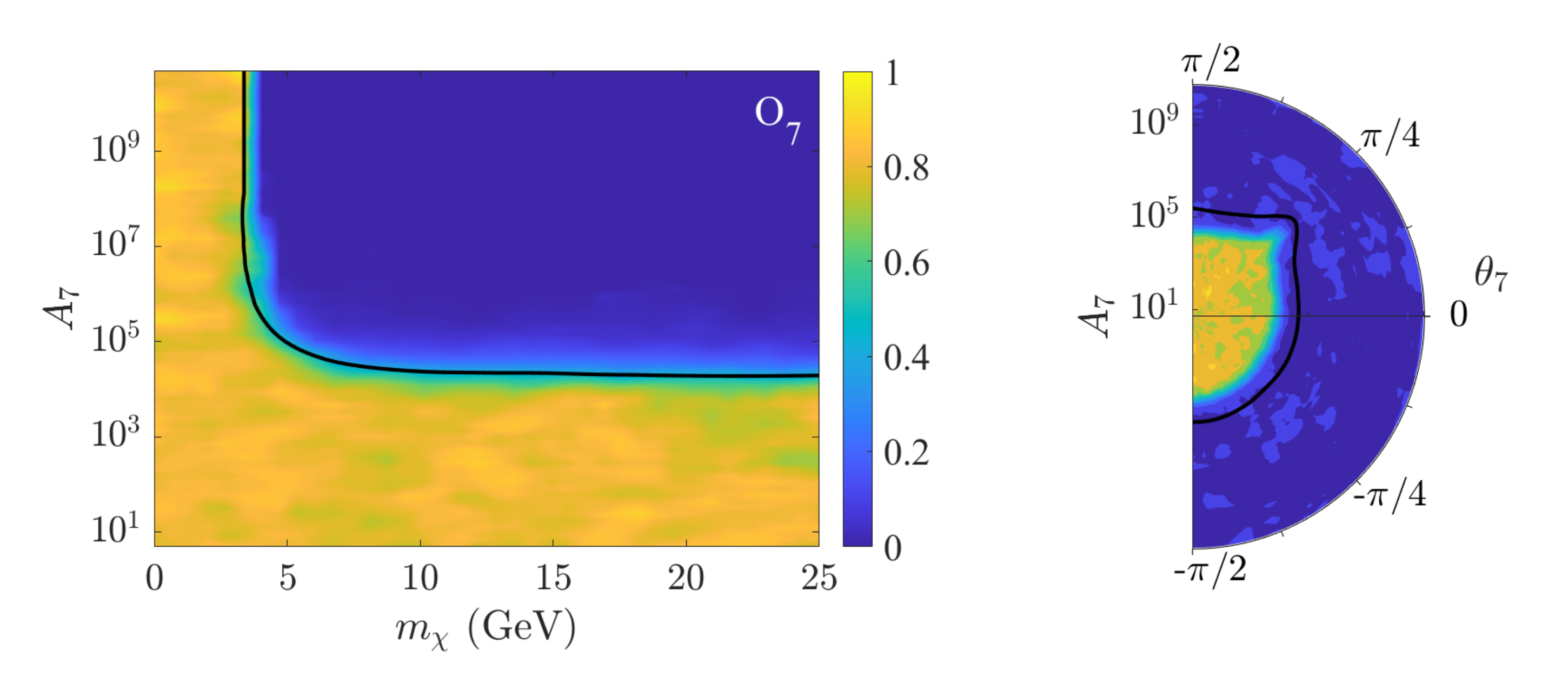}
   \includegraphics[width = 0.6\textwidth]{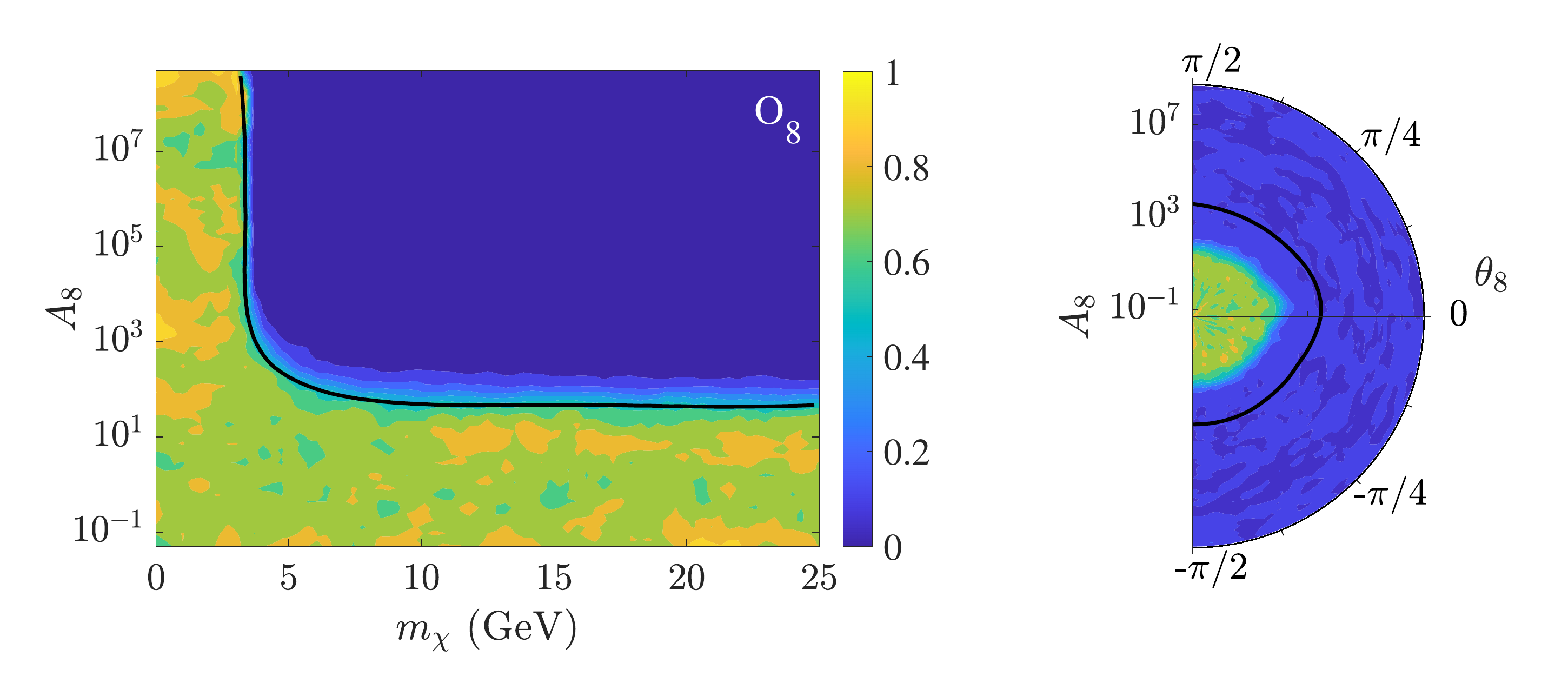}
   \includegraphics[width = 0.6\textwidth]{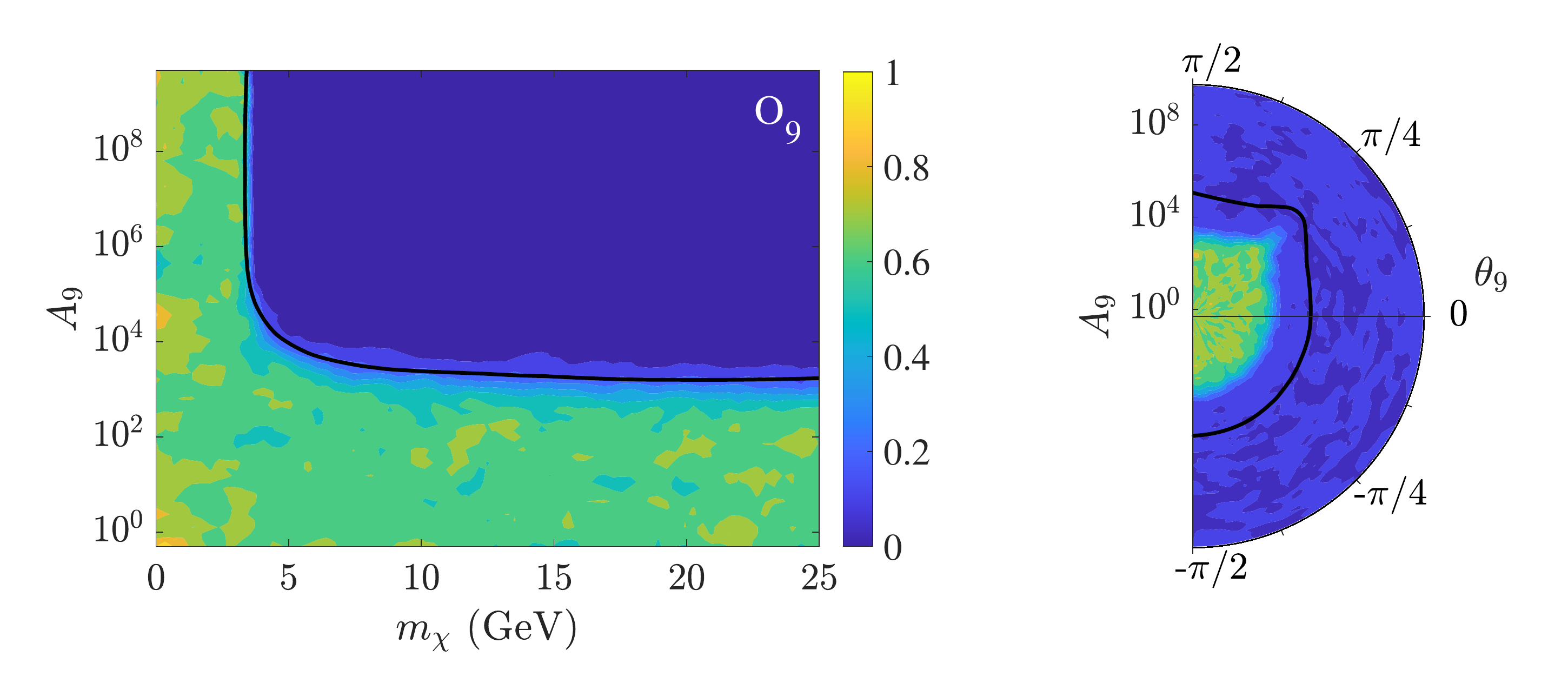}
   \caption{2D marginalized posterior distributions for the polar parameters ($m_\chi$, $A_i$, and $\theta_i$) of each EFT operator ($\mathcal{O}_6$ - $\mathcal{O}_9$) under the two-parameter background model. The respective 95\% credibility contours are shown in black.}
   \label{fig:2D_likelihoods_B}
\end{figure*}

\begin{figure*}[ht!]
   \centering
   \includegraphics[width = 0.6\textwidth]{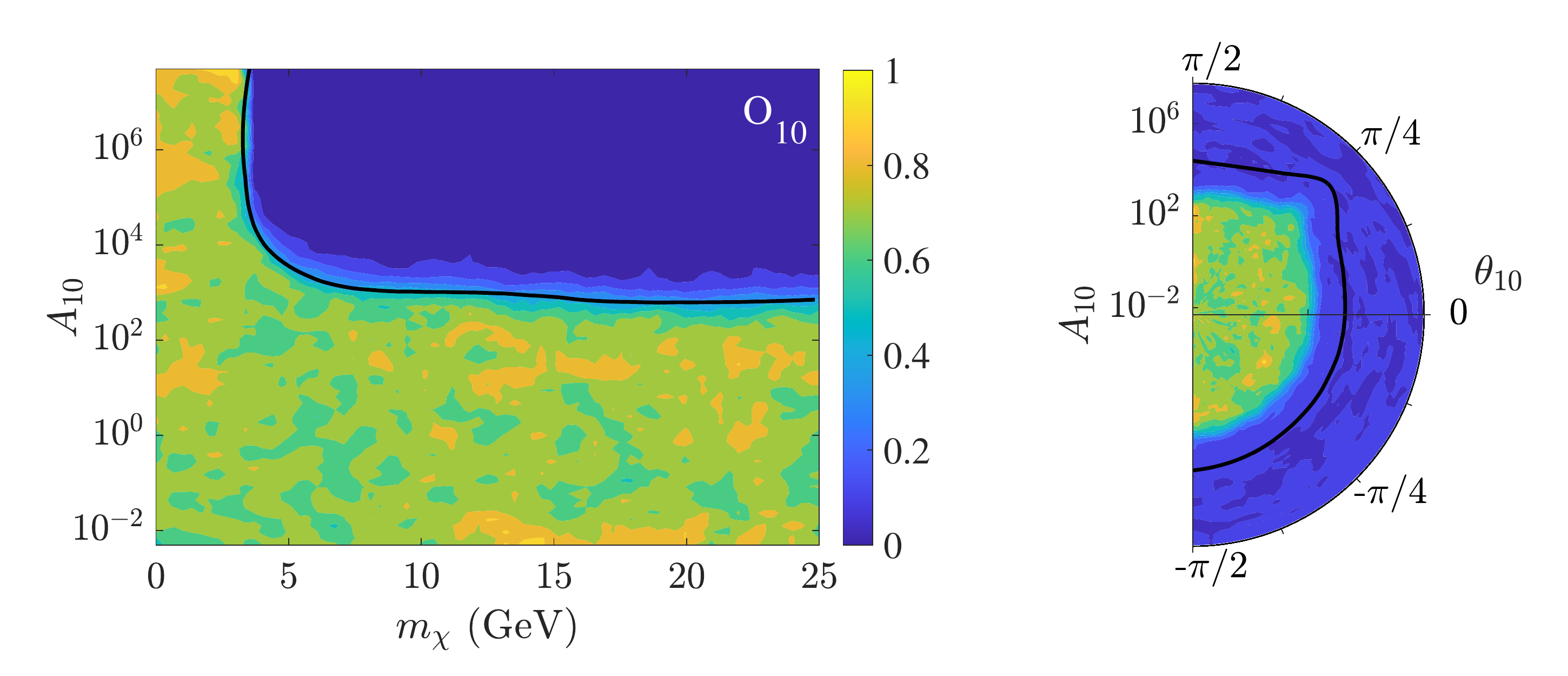}
   \includegraphics[width = 0.6\textwidth]{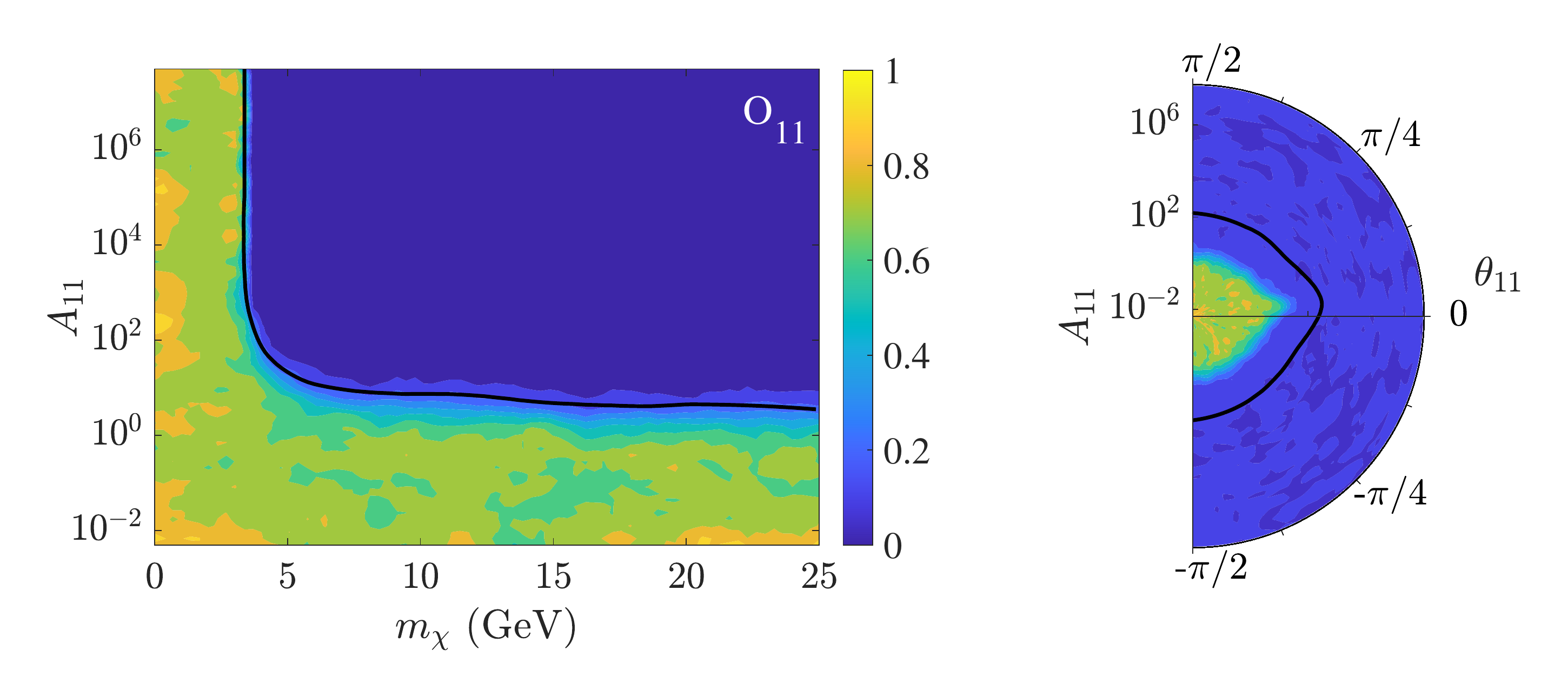}
   \caption{2D marginalized posterior distributions for the polar parameters ($m_\chi$, $A_i$, and $\theta_i$) of each EFT operator ($\mathcal{O}_{10}$ and $\mathcal{O}_{11}$) under the two-parameter background model. The respective 95\% credibility contours are shown in black.}
   \label{fig:2D_likelihoods_C}
\end{figure*}

The single operator models for the two-parameter background model analysis result in 5D likelihoods.
The 5D operator posterior distributions can be marginalized into three 2D posterior distributions on WIMP parameters and one 2D background posterior distribution.
Fig.~\ref{fig:2D_likelihoods_A}\,--\,\ref{fig:2D_likelihoods_C} show the resulting 2D WIMP parameter posterior distributions in the mass-amplitude plane and in the amplitude-phase plane for the prior choices of a flat mass prior and a flat $\log(A_i)$ prior (cf.~Eq.~(\ref{eq:flat_logA})).
Each marginalized posterior distribution is shown with 95\% Bayesian credibility contours.
In order to give an idea of which parameter regions are least and most favored, each marginalized posterior distribution is portrayed with colors varying from dark blue through yellow for the lowest through the highest probability.
In the mass-amplitude plane (left plot for each operator), all operators show contours whose shapes resemble that of the Optimum Interval upper limits given in a standard CDMS analysis (cf.~Fig.~\ref{fig:OI_comparison}) \cite{Agnese:2013jaa,Agnese:2015nto,Agnese:2017jvy,Agnese:2016cpb,Agnese:2013rvf,Agnese:2015ywx}. 
This is not unexpected, given that the Bayesian evidences show that the CDMSlite Run~2 data is consistent with the expected backgrounds. 
Although we could have also marginalised in the amplitude to show a mass-phase plane, most of these operators have flat posterior distributions below the 95\% credibility contour, and in the mass-phase plane the posterior distributions are nearly independent of phase.

\begin{figure*}[ht!]
   \centering
   \includegraphics[width = 0.28\textwidth]{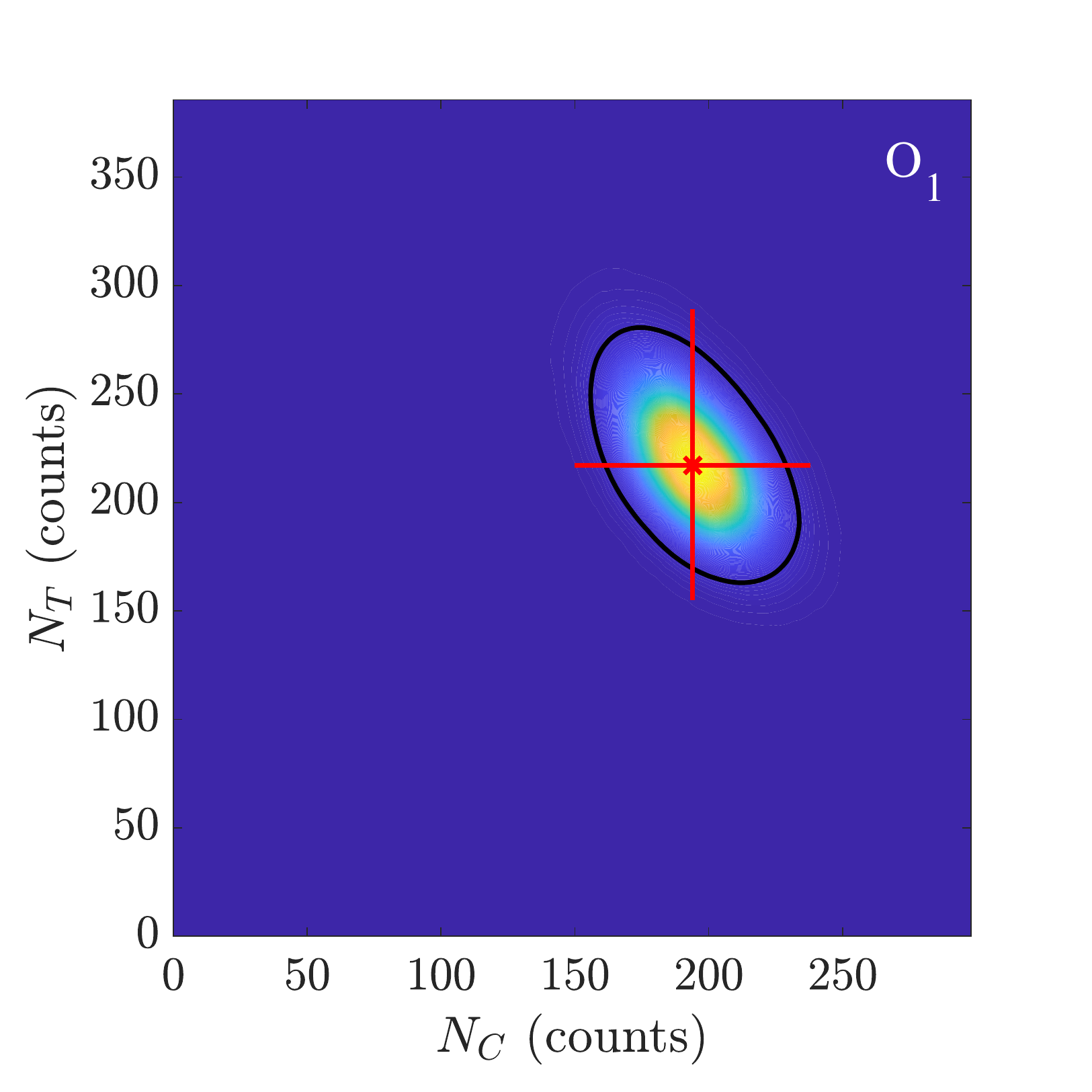}
   \includegraphics[width = 0.28\textwidth]{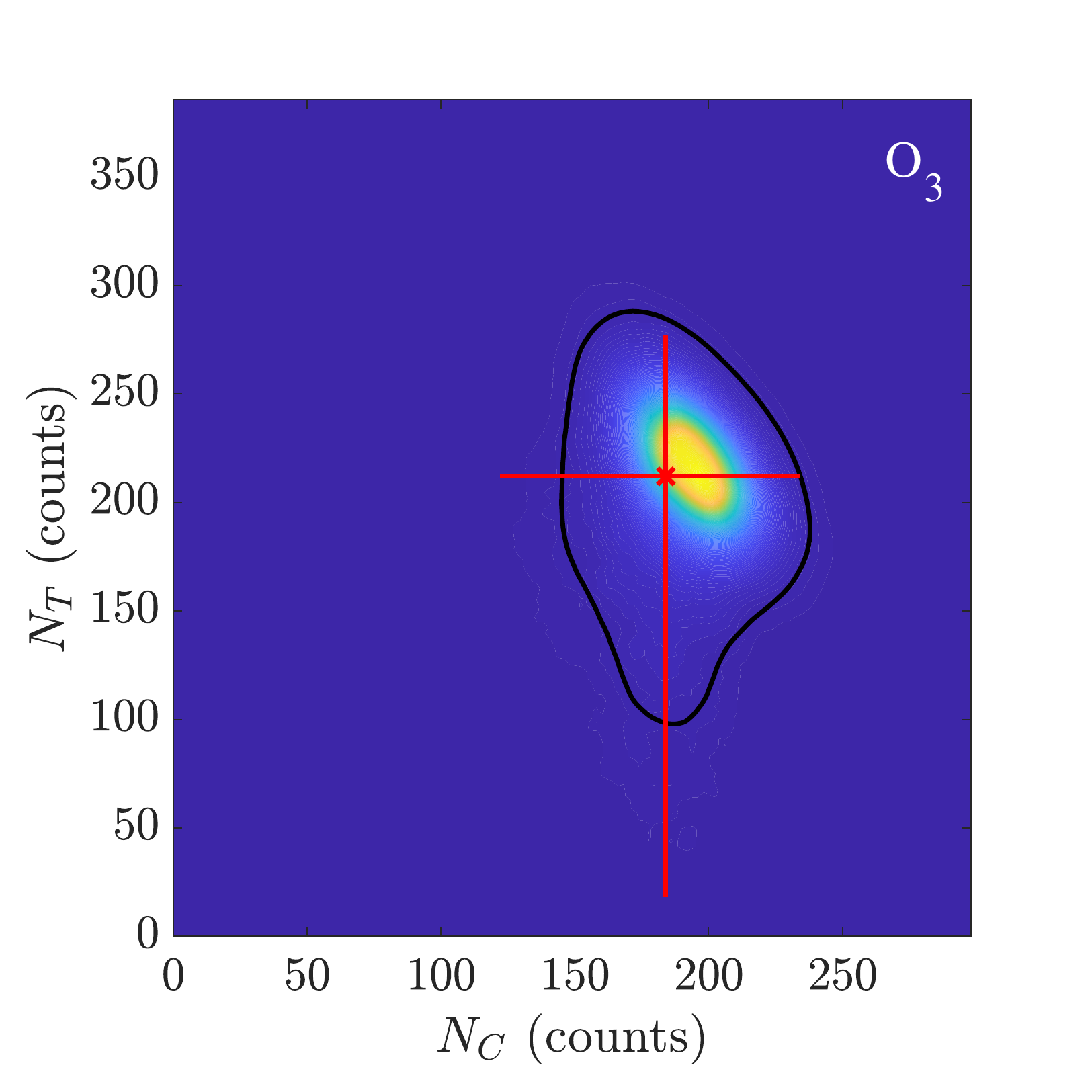}
   \includegraphics[width = 0.28\textwidth]{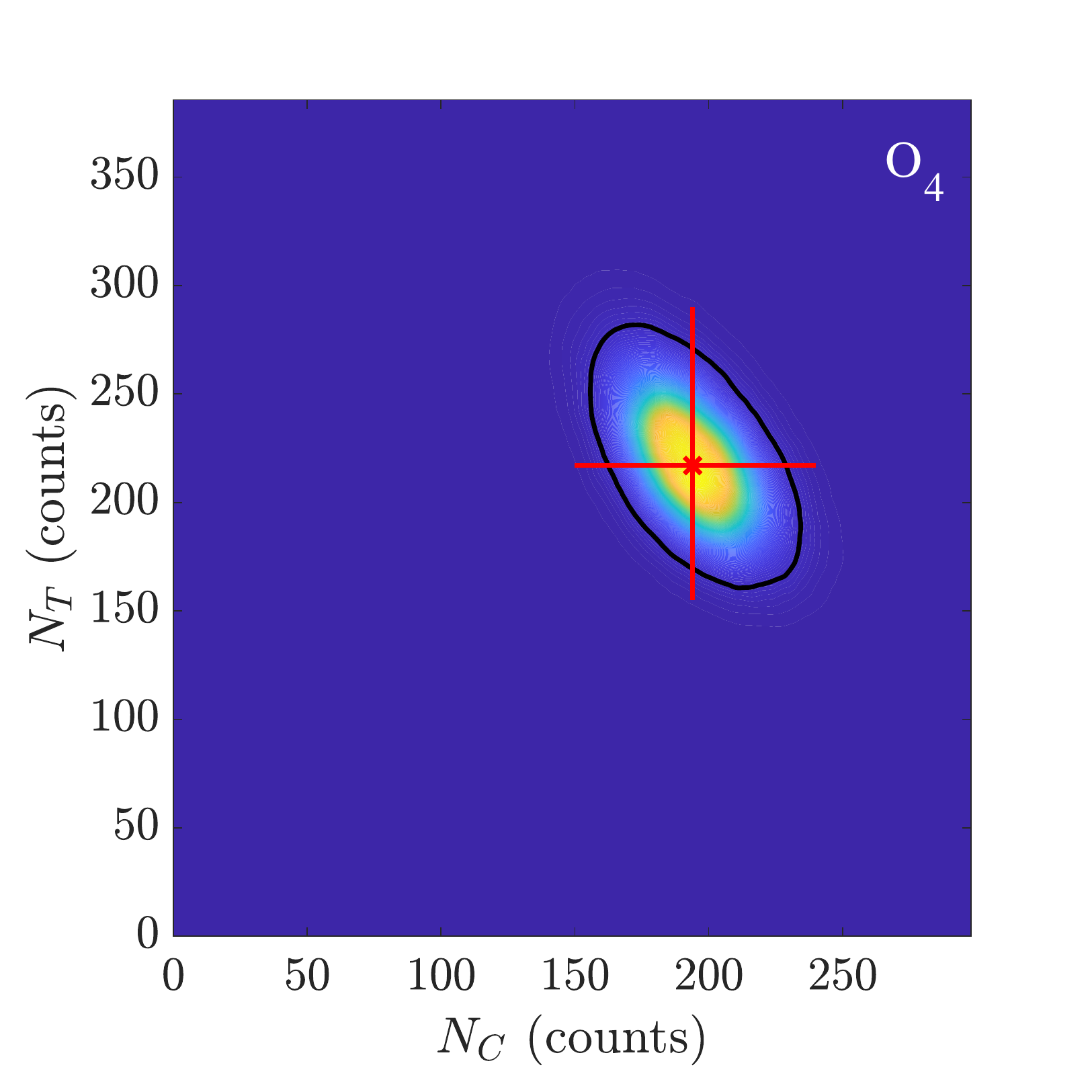}
   \includegraphics[width = 0.28\textwidth]{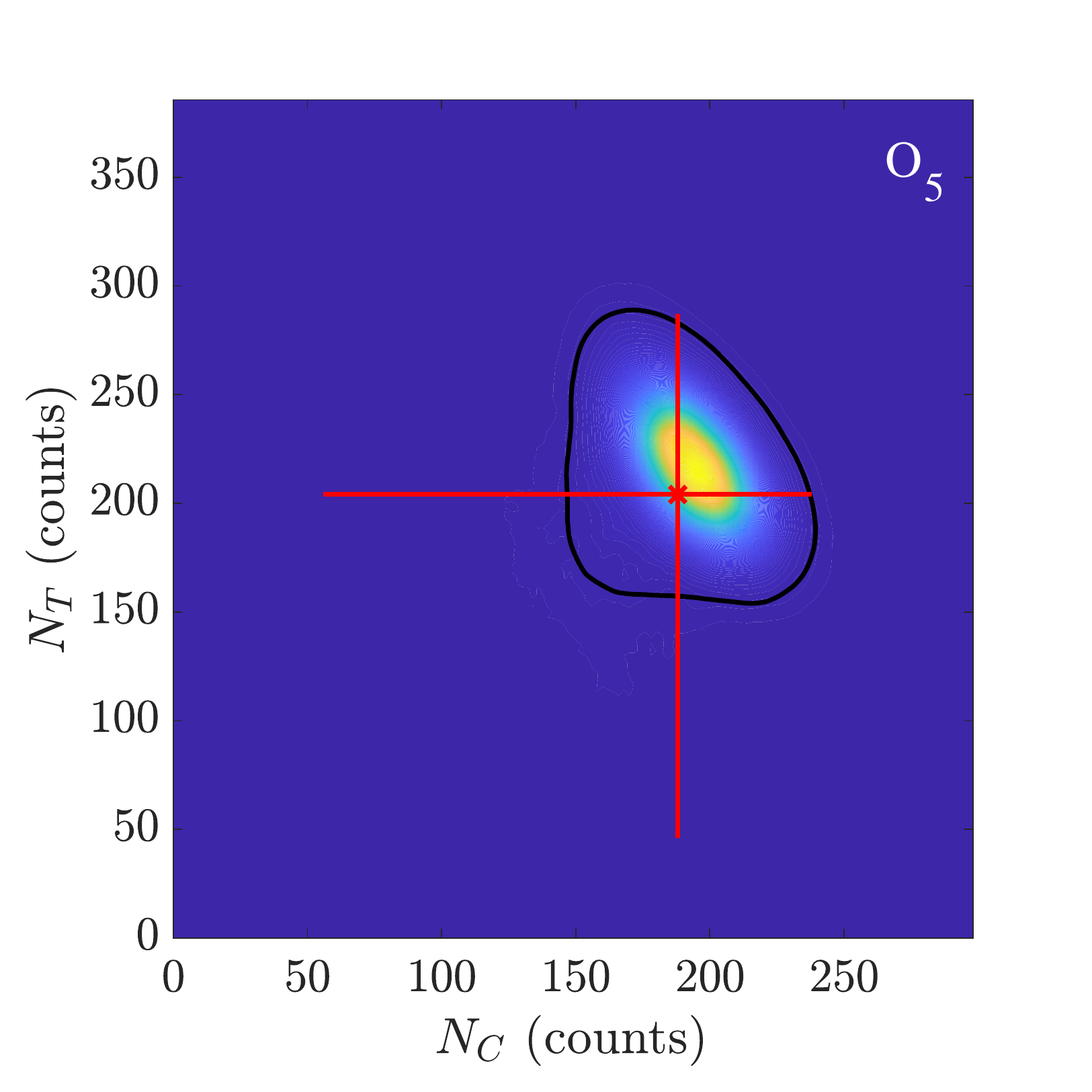}
   \includegraphics[width = 0.28\textwidth]{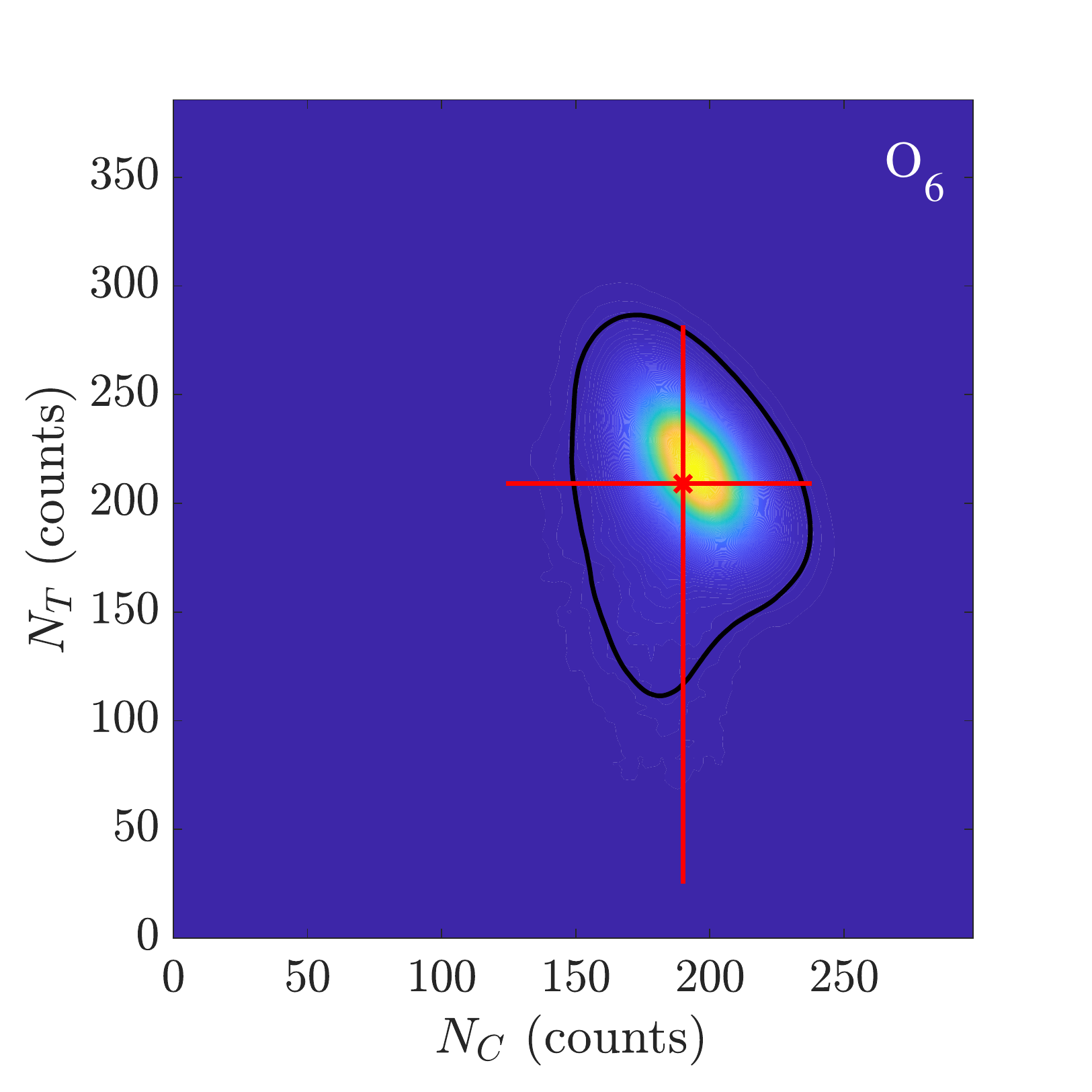}
   \includegraphics[width = 0.28\textwidth]{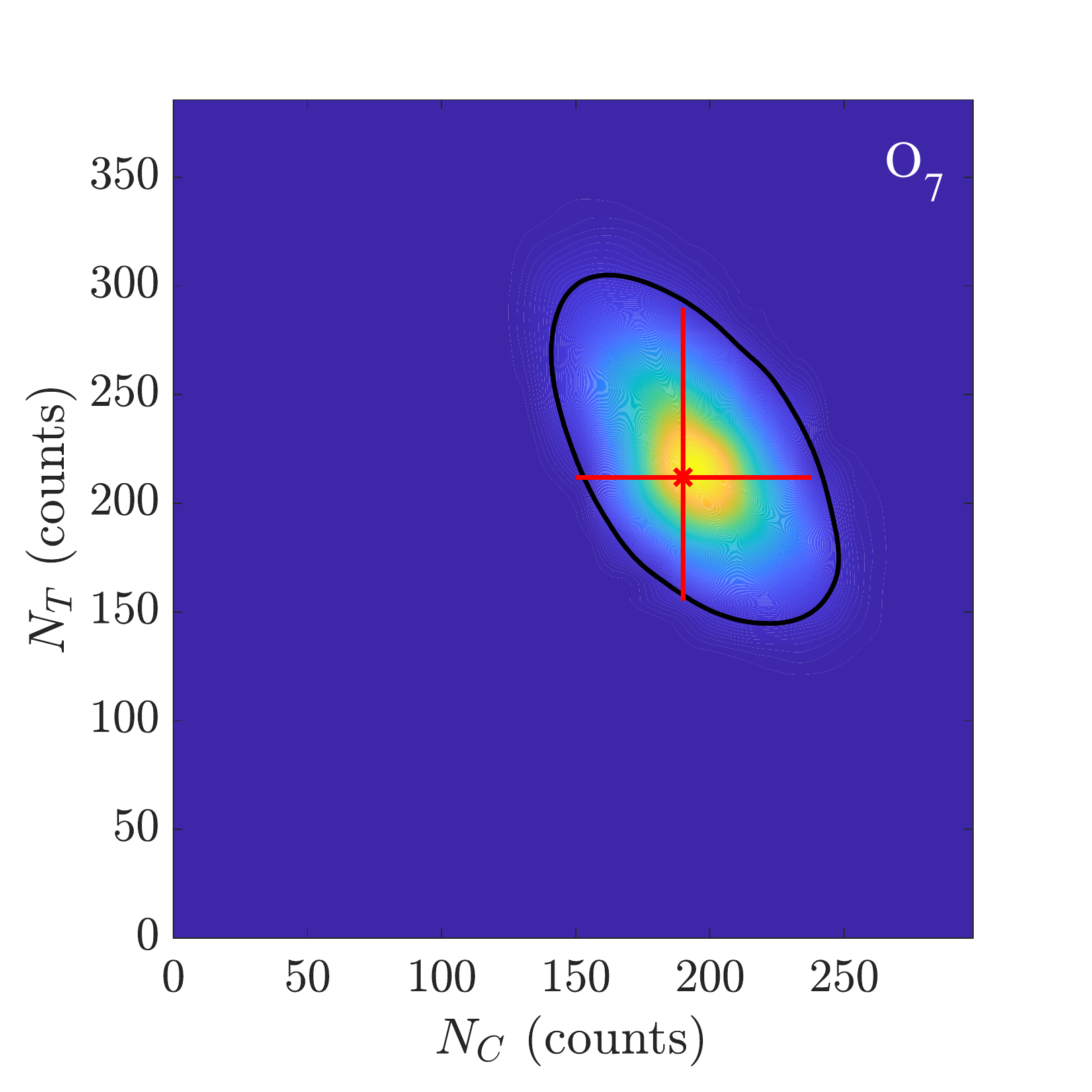}
   \includegraphics[width = 0.28\textwidth]{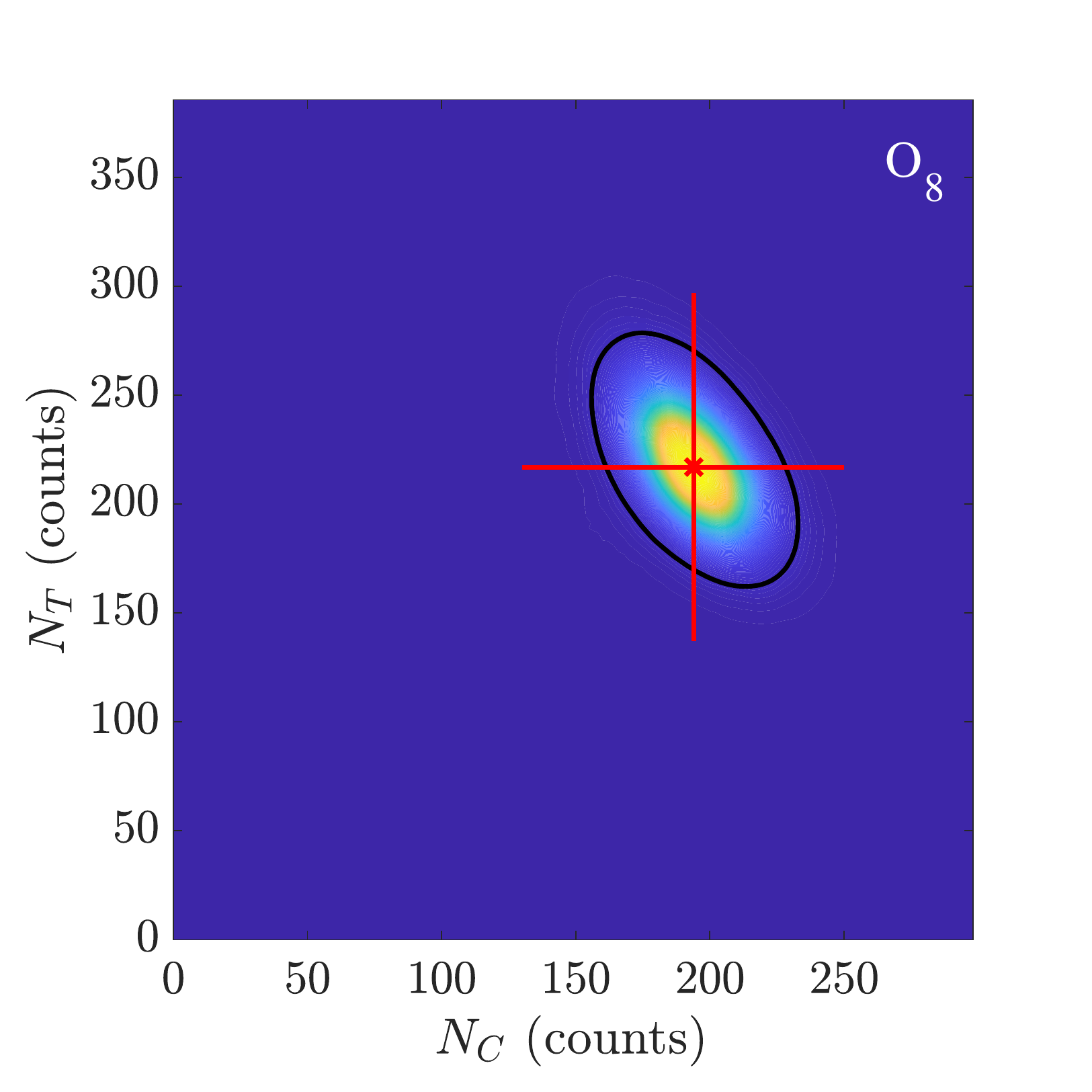}
   \includegraphics[width = 0.28\textwidth]{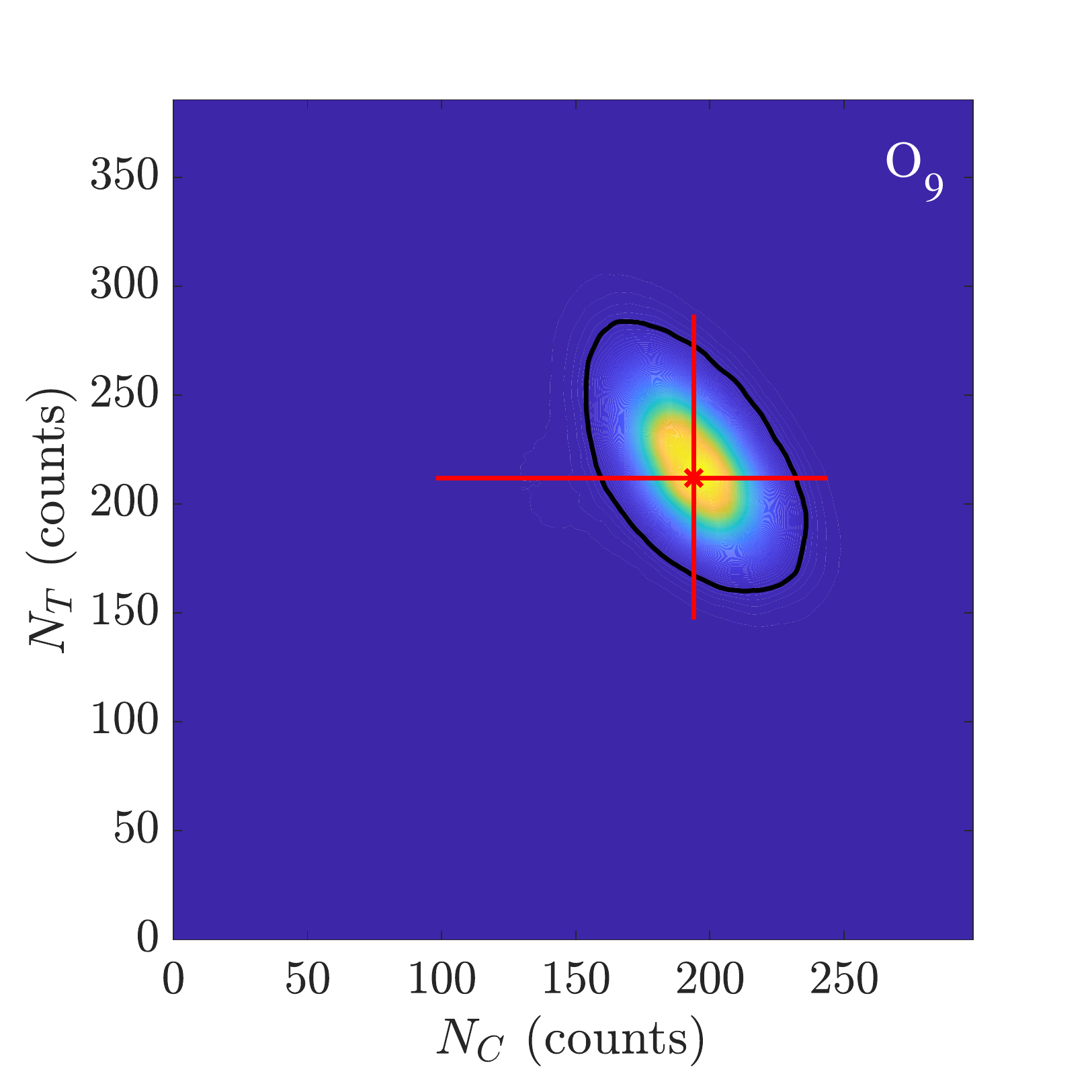}
   \includegraphics[width = 0.28\textwidth]{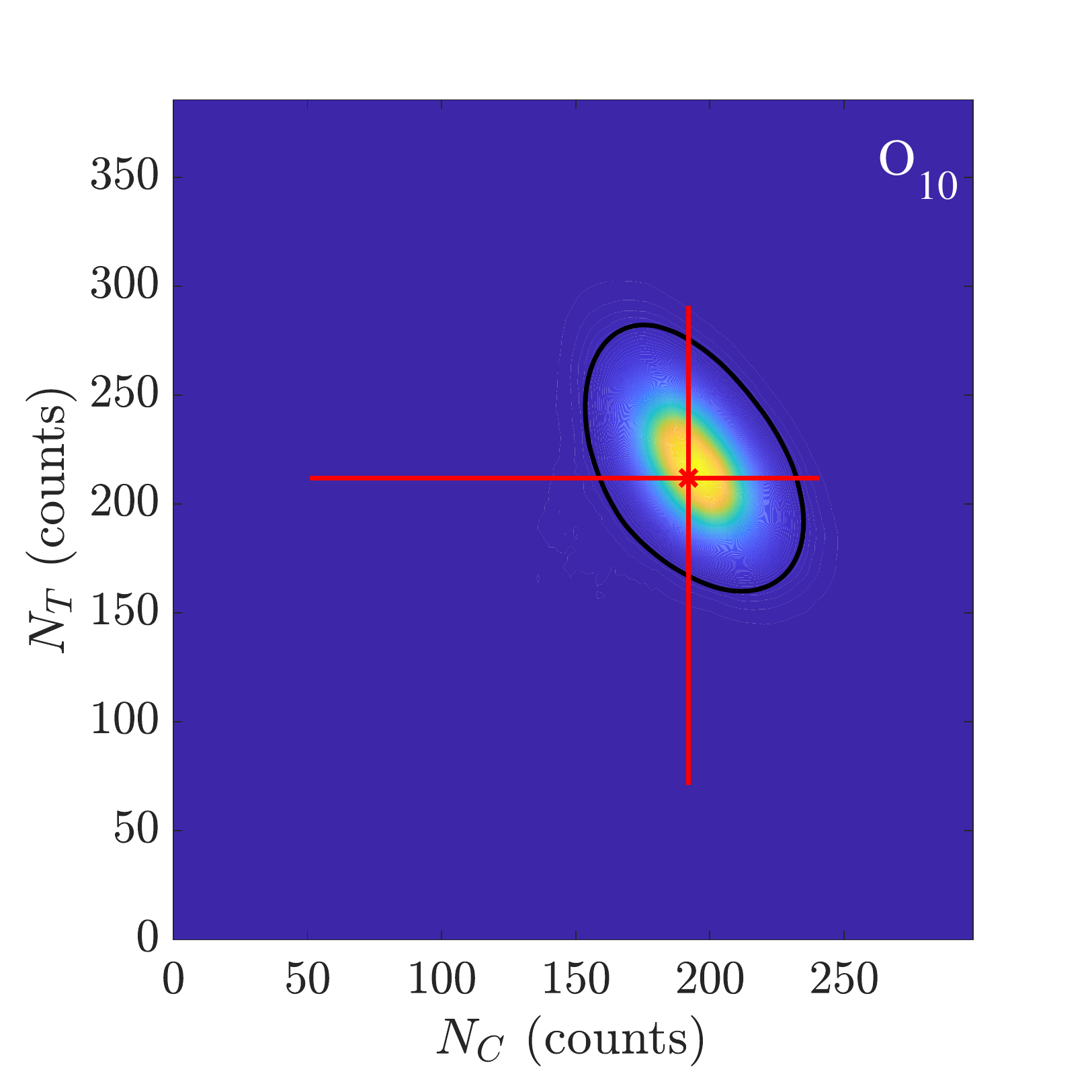}
   \includegraphics[width = 0.30\textwidth]{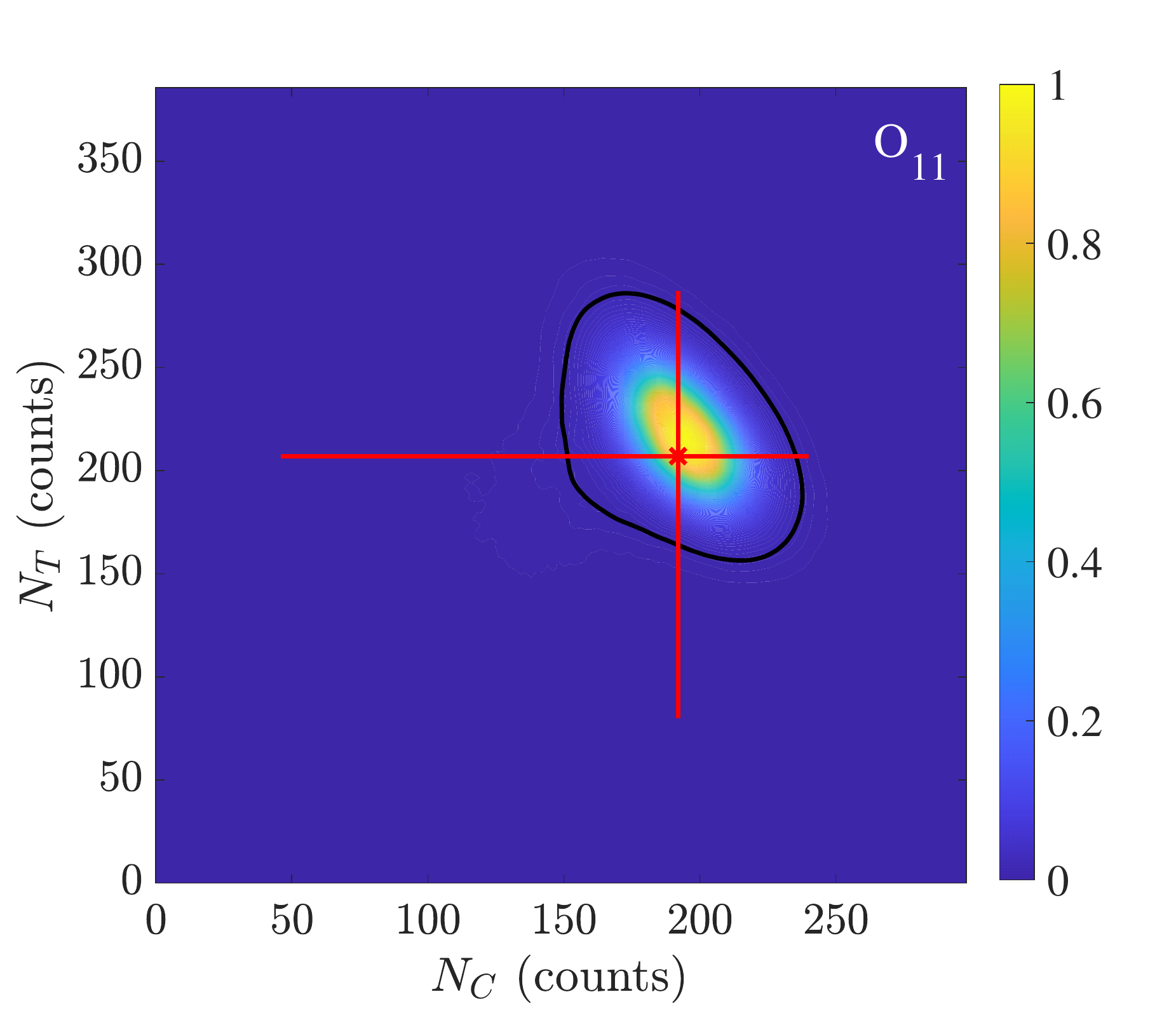}
   \caption{2D marginalized posterior distributions for the Compton and tritium backgrounds for each single EFT operator. The respective 95\% credibility contours are shown in black. The point of maximum likelihood is indicated as a red cross with the corresponding $2\sigma$ uncertainty calculated from the 1D marginalized likelihoods of each background source separately.}
   \label{fig:2D_likelihoods_bgk}
\end{figure*}

The amplitude-phase plane (displayed in polar coordinates) shows a tendency to favor lower amplitudes with spikes into higher amplitudes at particular angles.
A spike in the 95\% credibility contour represents directions in which the limit on the amplitude is weaker because of destructive interference in the phase of that operator for Ge.
The maximum interference phase is where the isoscalar and isovector components of the coupling coefficients have the largest destructive interference.
Destructive interference causes the amplitudes of the theoretical WIMP signal to decrease, meaning it takes a larger amplitude to get to the same number of events as for other phases.
Because the maximum interference phase can be different for different target materials, combining the CDMSlite Run 2 data with data from another experiment (such as the future SuperCDMS HV Si data \cite{Agnese:2016cpb}) can set stronger limits even if there is no WIMP detection \cite{Rogers:2016jrx,Bertone:2007xj,Cerdeno:2013gqa,Pato:2010zk}.

The SD operators $\mathcal{O}_4$, $\mathcal{O}_6$, $\mathcal{O}_7$, $\mathcal{O}_9$ and $\mathcal{O}_{10}$ and the SI operator $\mathcal{O}_5$ show destructive interference at $\theta = \pi/4$, which represents proton-coupling.
The operators $\mathcal{O}_1$, $\mathcal{O}_8$ and $\mathcal{O}_{11}$ show a slight weakness towards isovector-couplings ($\theta=0$).

The background likelihood for each operator is also two-dimensional.
The corresponding 2D posterior distributions are shown in Fig.~\ref{fig:2D_likelihoods_bgk}, similarly to the two-parameter background-only posterior distribution in Fig.~\ref{fig:twoparameter_bkg_only}.
Some of the operators ($\mathcal{O}_1$, $\mathcal{O}_4$, $\mathcal{O}_7$, $\mathcal{O}_8$ and $\mathcal{O}_9$) have posterior distributions that look near-Gaussian for both dimensions.
Operators $\mathcal{O}_3$ and $\mathcal{O}_6$ have near-Gaussian distributions for the Compton background, but the tritium posterior distributions have long tails into low background counts.
The posterior distributions of the remaining operators ($\mathcal{O}_5$, $\mathcal{O}_{10}$ and $\mathcal{O}_{11}$) have long tails into low background counts for both sources considered. 

Even with some variability in shape, the point of maximum likelihood for each background source is fairly consistent from operator to operator.
For the Compton background, this peak of highest likelihood ranges from 184 to 194 counts with an average of 192$\pm$20 counts.
The background-only model predicts 168$\pm^{56}_{48}$ counts for the Compton background.
The tritium background peak ranges from 200 to 217 counts with an average of 214$\pm$22 counts, compared to 252$\pm$57 events predicted by the background-only model.
When including the EFT operators in the fit, the background models predict systematically more Compton events and fewer tritium events, which is consistent with some of the EFT operators producing $\beta$-decay-like recoil spectrum shapes.
However, the statistical significance of this trend is only on the level of $1\sigma$ considering the uncertainties.

\section{Conclusion} \label{sec:conclusions}
In this article we presented the first dark matter EFT analyses of CDMSlite data.
The techniques used to handle the background include removing energy bins around known activation peaks and modeling the shapes of the remaining backgrounds (Compton and tritium) with an analytic two-parameter model.
Posterior distributions were calculated for combinations of different background models (both backgrounds included separately, each background included on its own, and no backgrounds) and each EFT operator.

The Bayesian evidences of these likelihoods show that the CDMSlite Run 2 data are well described by the Compton and tritium backgrounds, and that the fit does not benefit from including any of the EFT operators.
Using the two-parameter background-only posterior distribution, the Compton background is estimated to be 168$\pm^{56}_{48}$ events, and the background from tritium to be 252$\pm57$ events, consistent with a prior dedicated analysis \cite{Agnese:2019app}.
However, some of the EFT operators do fit the data better than others, with operators $\mathcal{O}_{1}$, $\mathcal{O}_{4}$, $\mathcal{O}_{7}$ and $\mathcal{O}_{10}$ having a ratio of evidence to the background-only likelihood of greater than 0.45.
Having data consistent with background, 95\% Bayesian credibility upper limits were calculated in the mass-amplitude plane of each EFT operator.
These are similar to (but not directly comparable to) the cross-section versus WIMP-mass upper limits typically published by dark matter experiments.

\appendix

\section{Choice of Prior Probability Distributions}
\label{app:priors}

In order to include each parameter describing an EFT operator in the likelihood calculation, the prior probability distribution of each parameter must be determined.
For the WIMP mass, there are two options considered for prior probability distributions.
The first, a flat prior, is given by the equation
\begin{equation}
\text{Pr}(m_{\chi}) = \left\{ \begin{array}{l}
    {\frac{1}{m_\text{max}}, ~0\leq m_{\chi} \leq m_\text{max}} \\
    {0,\text{ otherwise}} \\
    \end{array} \right. ,
 \end{equation}
where $m_\text{max} = 45$\,GeV/$c^2$. This upper bound is higher than the bound used in the main analysis.
The main analysis uses $m_\text{max} = 25$\,GeV/$c^2$.
The second option is a flat-log prior with a lower bound of $m_\text{min} = 0.1$\,GeV/$c^2$ and is given by
\begin{eqnarray}
\text{Pr}(\log m_{\chi}) &=& \left\{ \begin{array}{l}
    {\frac{1}{\log(m_\text{max})-\log(m_\text{min})}} \\
    {0,\text{ otherwise}} \\
    \end{array} \right. \\
    & & \text{with } m_\text{min}\leq m_{\chi} \leq m_\text{max} \nonumber.
 \end{eqnarray}

\begin{figure*}[ht!]
   \centering
   \includegraphics[width = 0.28\textwidth]{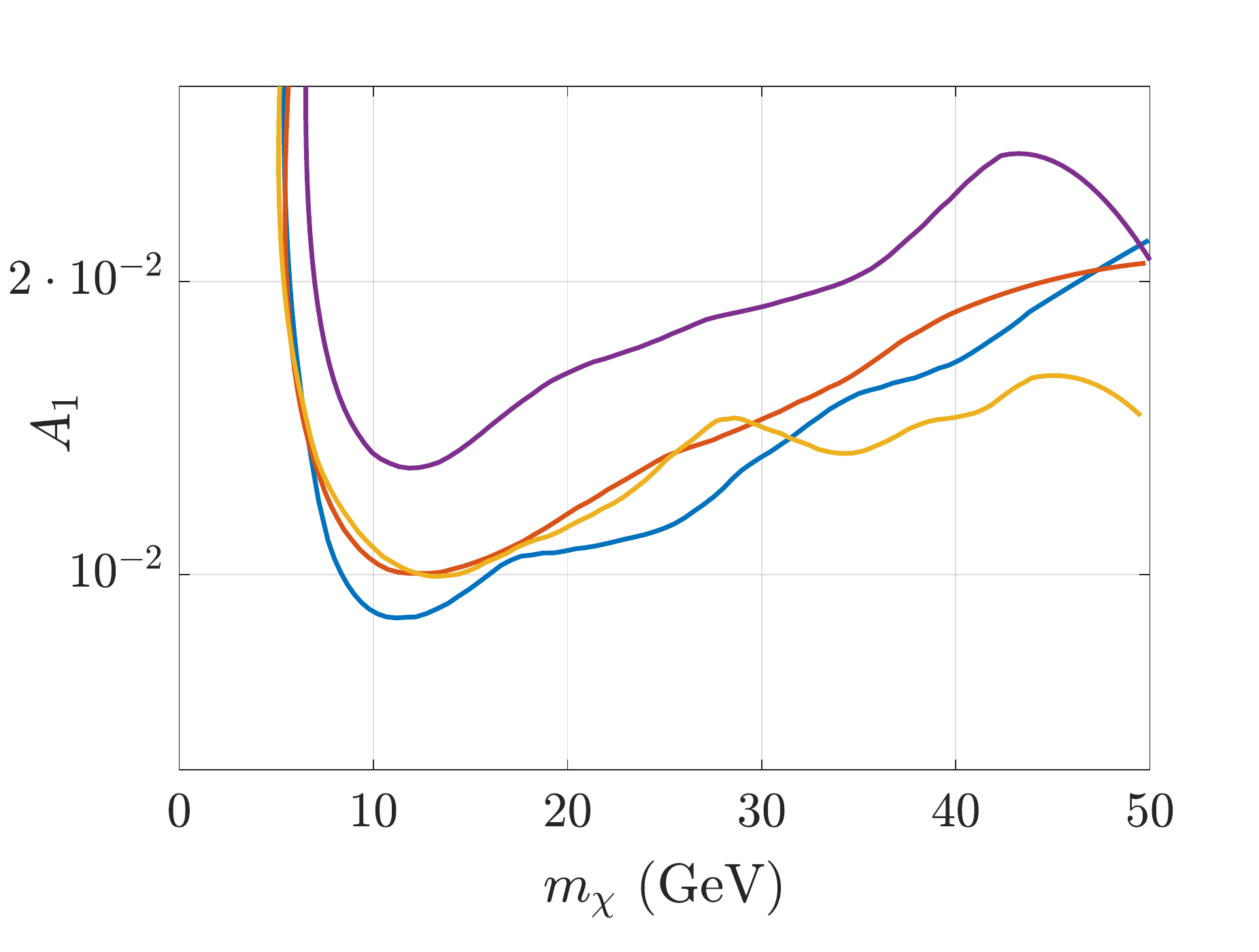}
   \includegraphics[width = 0.28\textwidth]{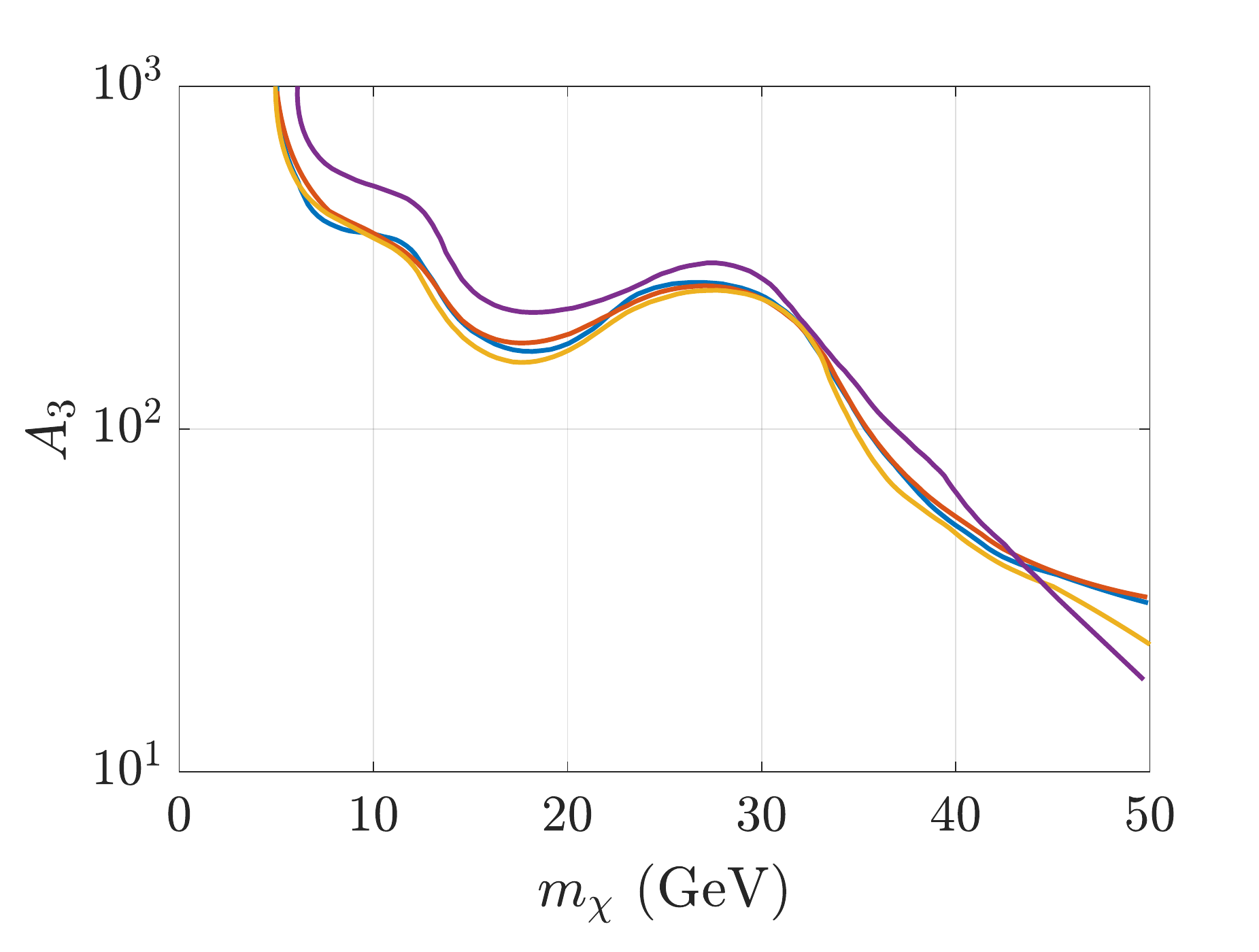}
   \includegraphics[width = 0.28\textwidth]{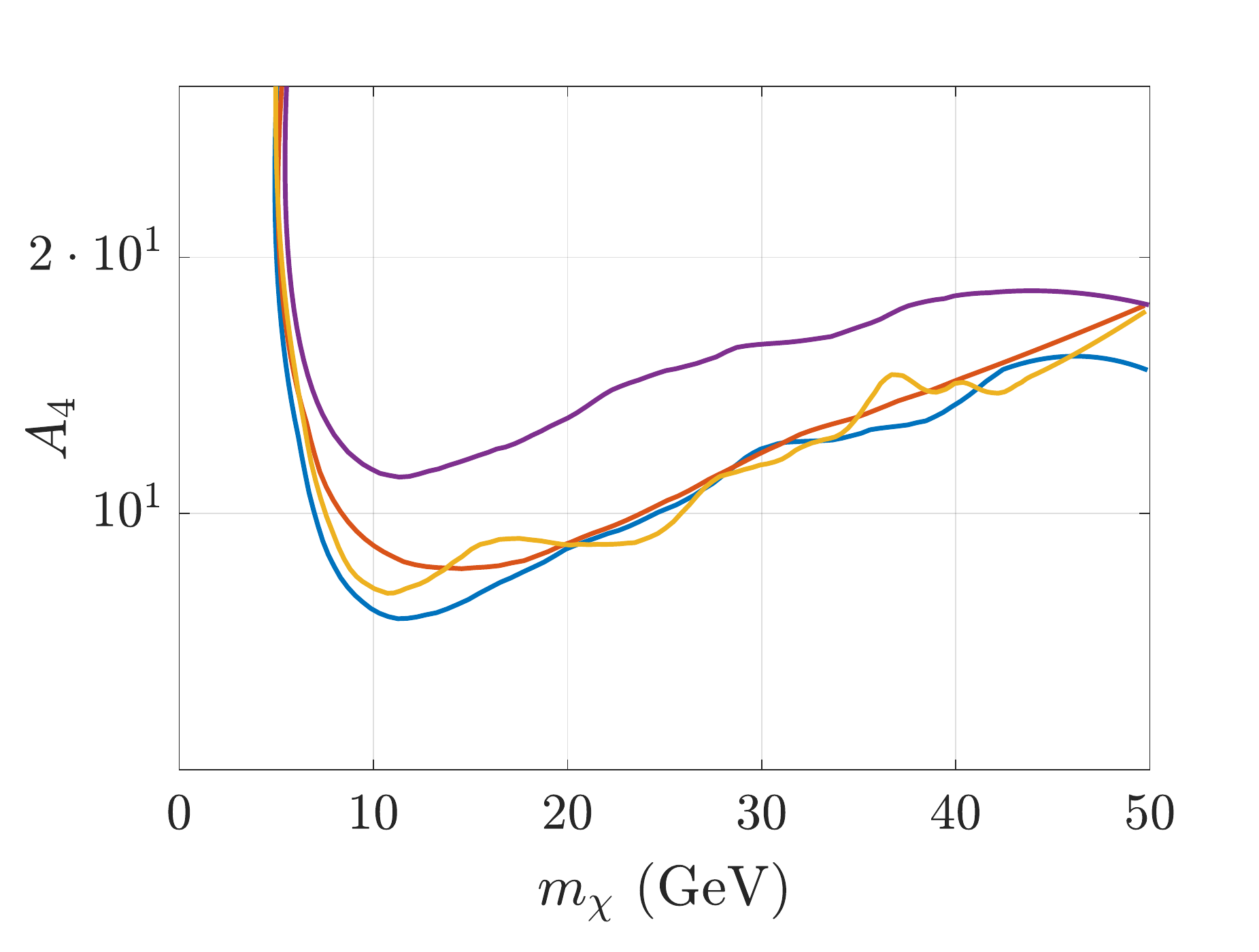}
   \includegraphics[width = 0.28\textwidth]{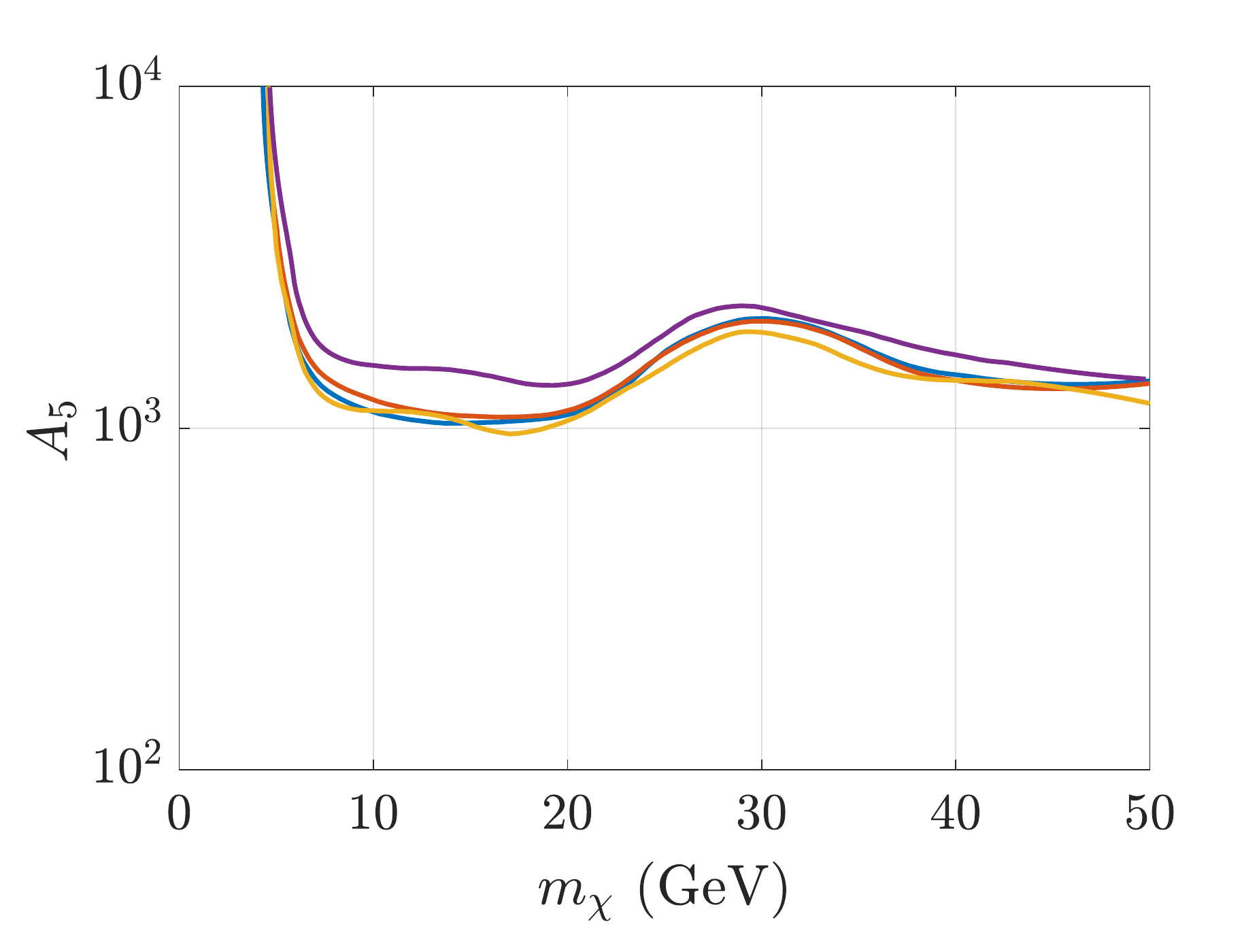}
   \includegraphics[width = 0.28\textwidth]{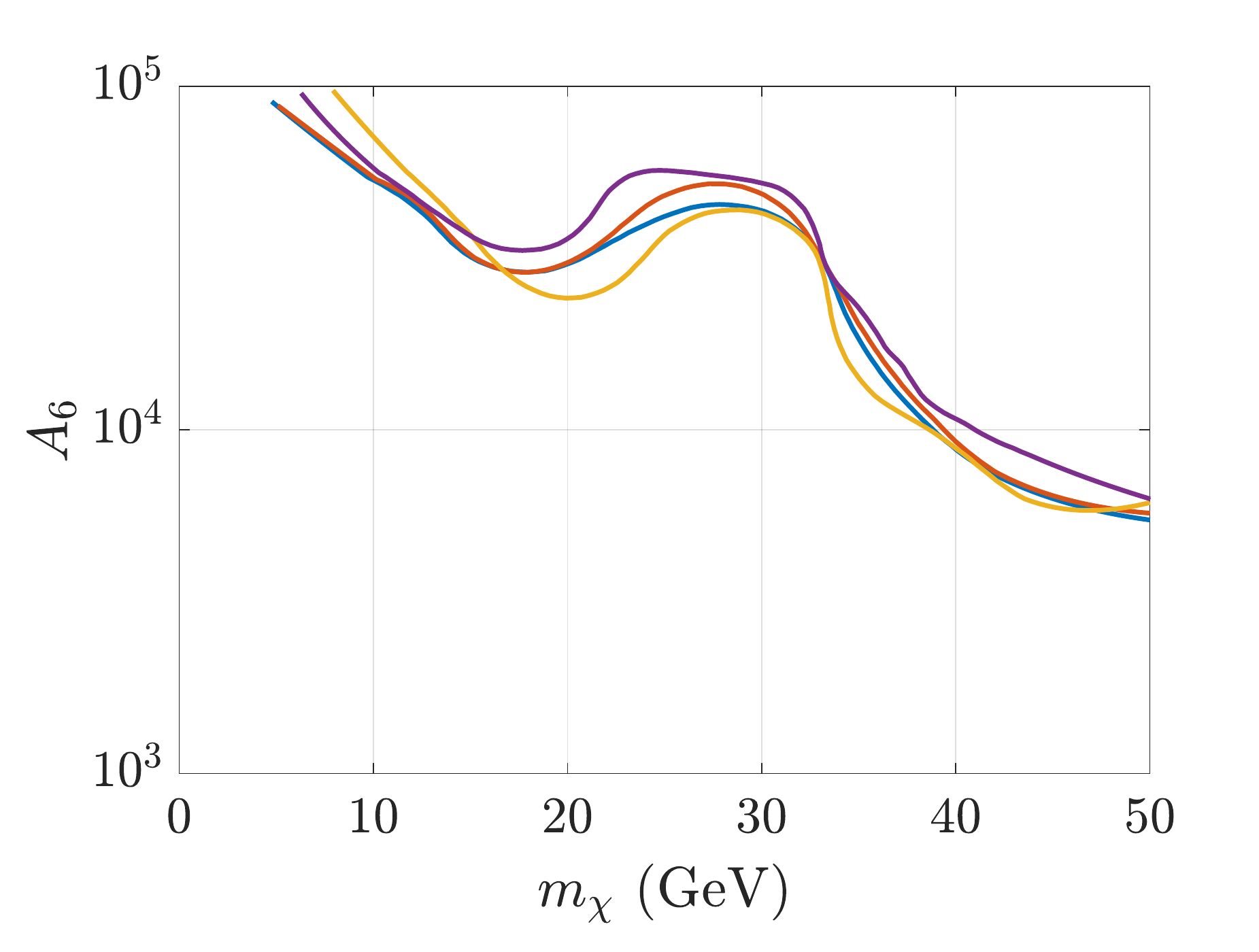}
   \includegraphics[width = 0.28\textwidth]{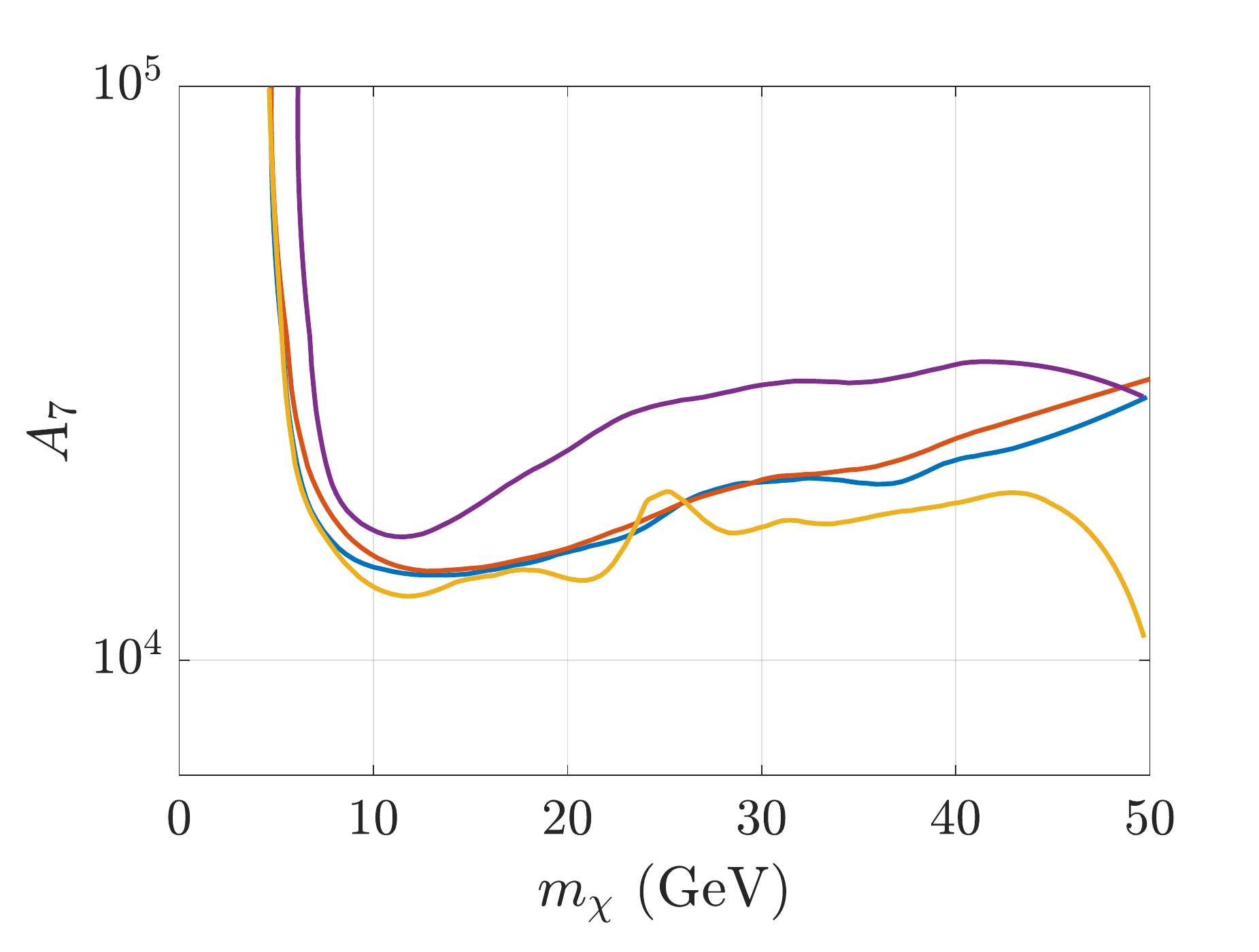}
   \includegraphics[width = 0.28\textwidth]{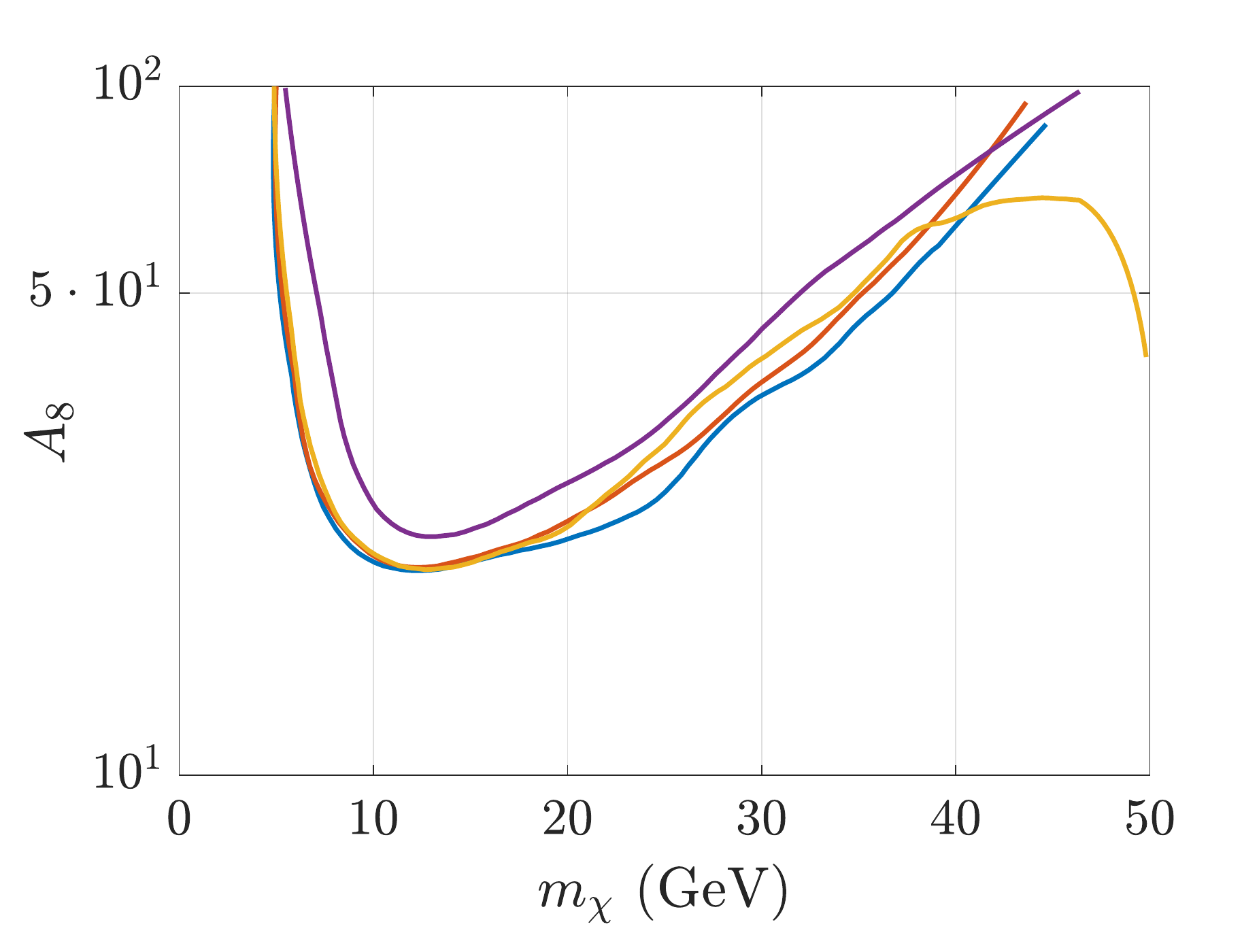}
   \includegraphics[width = 0.28\textwidth]{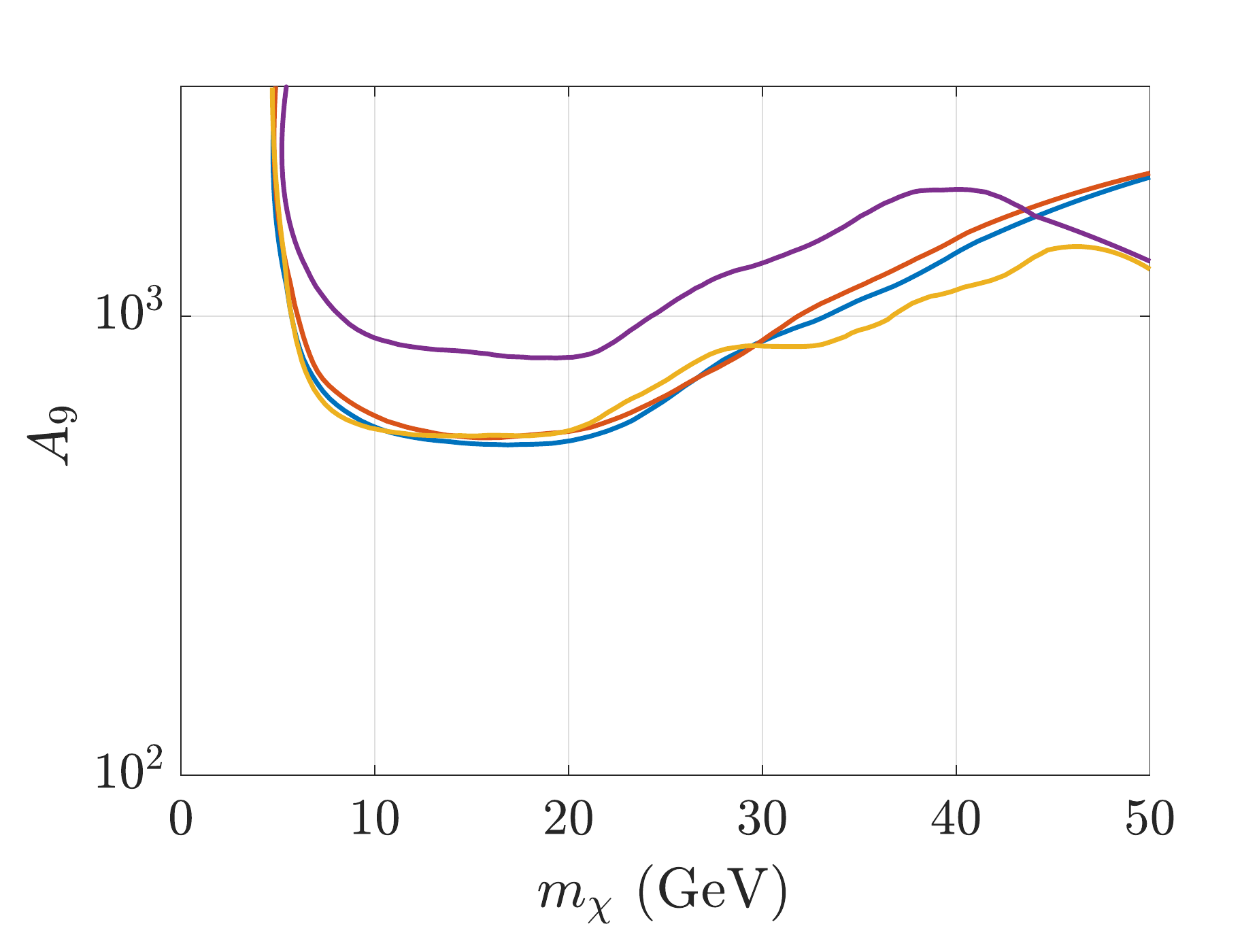}
   \includegraphics[width = 0.28\textwidth]{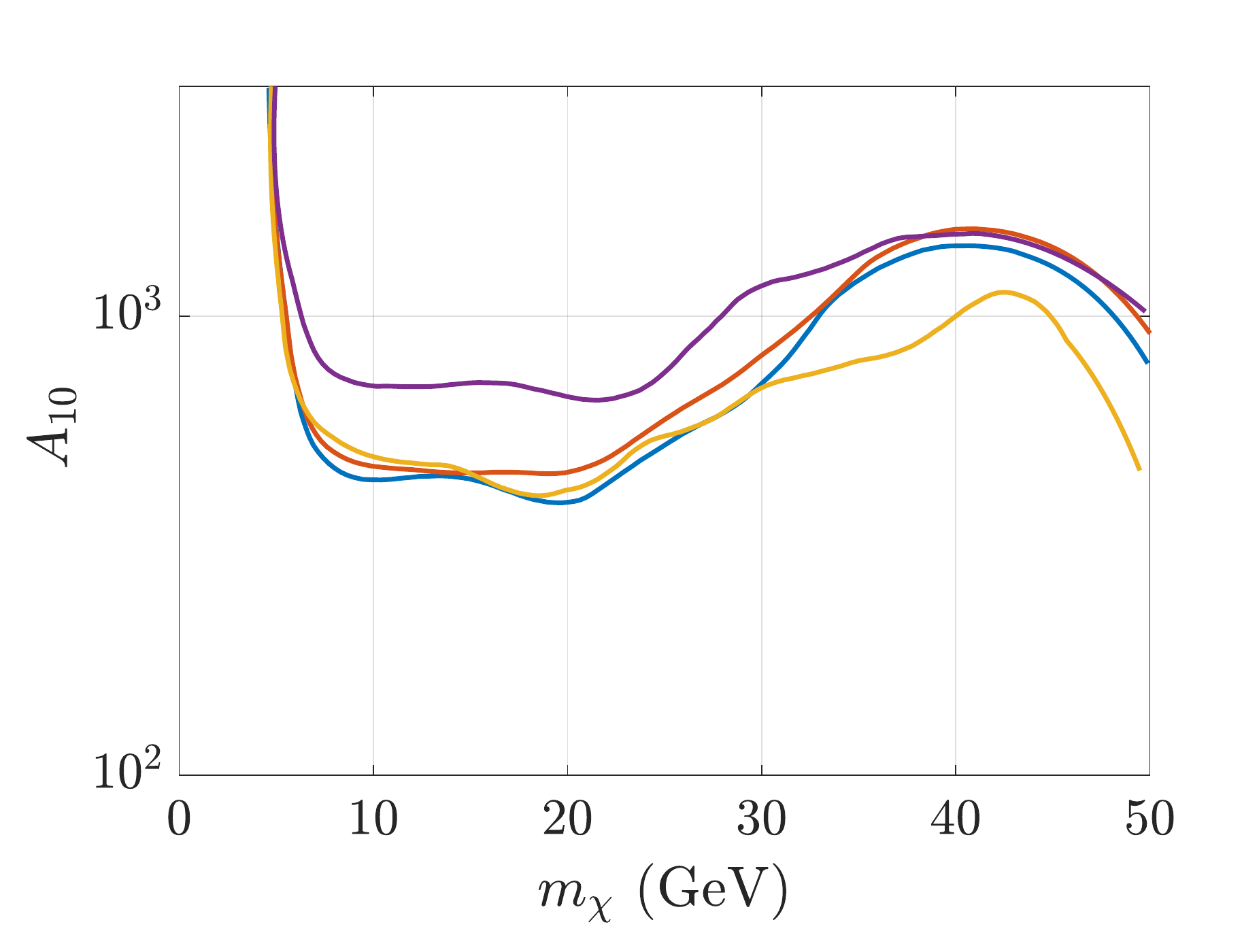}
   \includegraphics[width = 0.40\textwidth]{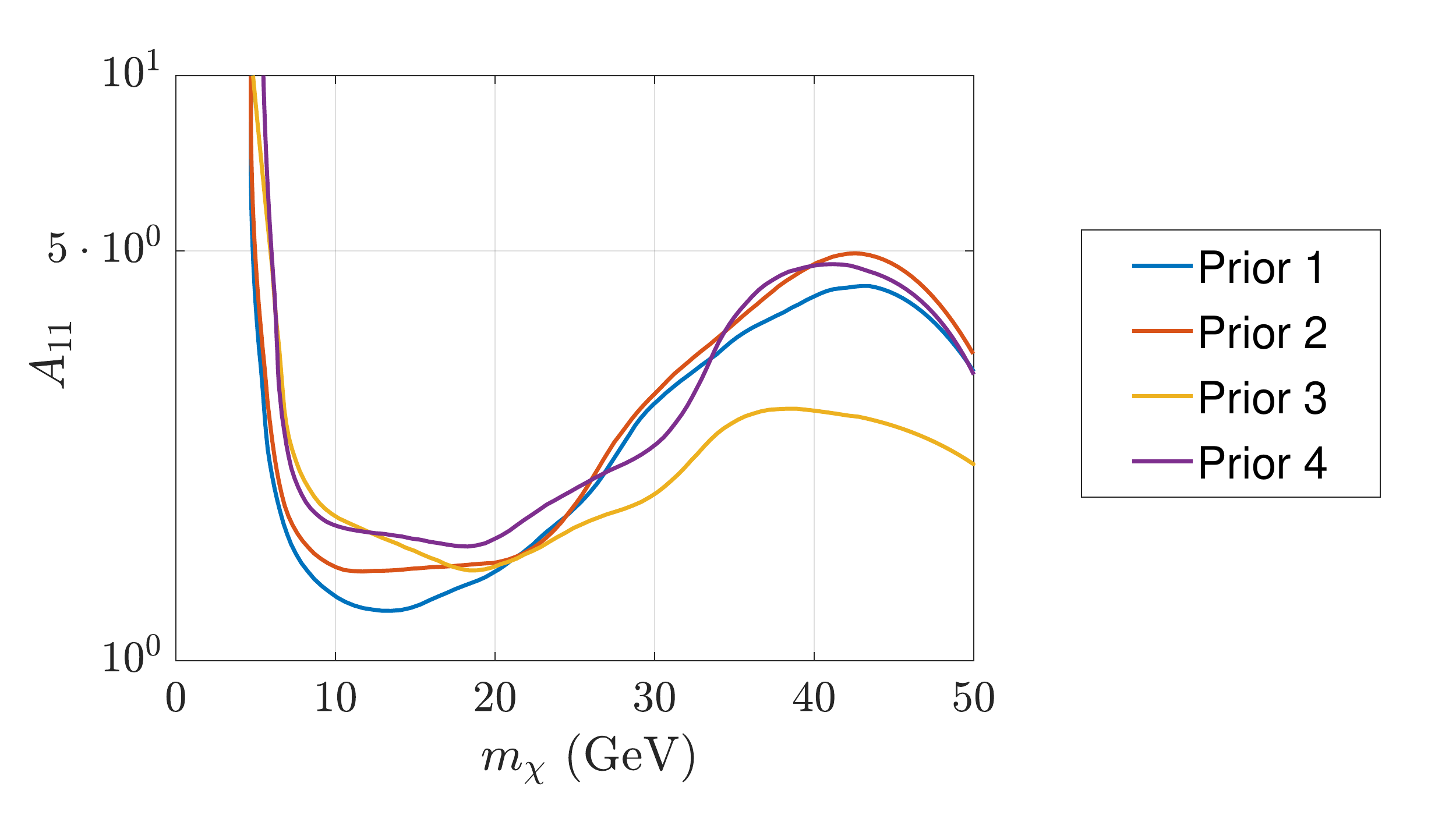}
   \caption{Comparison of 95\% credibility posterior contours on the coupling-coefficient amplitude and WIMP mass using each prior combination for each EFT operator.
   }
   \label{fig:prior_compare}
\end{figure*}

Similarly, there are two possible prior probability distribution choices for the coupling-coefficient amplitude.
The flat prior is given by
\begin{eqnarray}
\text{Pr}(A_i) &=& \left\{ \begin{array}{l}
    {\frac{1}{A_{i,\text{max}}-A_{i,\text{min}}}} \\
    {0,\text{ otherwise}} \\
    \end{array} \right. \\
     & & \text{with } A_{i,\text{min}} \leq A_i \leq A_{i,\text{max}}, \nonumber
 \end{eqnarray}
and the flat-log prior is given by 
\begin{eqnarray}
\text{Pr}(\log A_i) &=& \left\{ \begin{array}{l}
    {\frac{1}{\log(A_{i,\text{max}})-\log(A_{i,\text{min}})}} \\
    {0,\text{ otherwise}} \\
    \end{array} \right. \\
     & & \text{with } A_{i,\text{min}} \leq A_i \leq A_{i,\text{max}}, \nonumber
 \end{eqnarray}
where $A_{i,\text{max}}$ and $A_{i,\text{min}}$ were chosen for each operator to span roughly 13 orders of magnitude around a low number of expected events that still allows for the detected spectrum to be a WIMP signal.
In this way, the size of the parameter space for different operators remains the same, which removes the need for additional normalization before comparing the resulting Bayesian evidences. 
Because the amplitude is calculated over such a large number of decades, the flat prior cannot sample the full likelihood in a meaningful way.
Therefore, in order to compare the flat prior with the flat-log prior, the limits ($A_{i,\text{max}}$ and $A_{i,\text{min}}$) were modified to cover only two orders of magnitude around the 95\% credibility contour for the posterior distribution.

Using these sets of prior probability distributions results in four possible combinations of priors:
\begin{itemize}
\item Prior 1: flat $m_\chi$ and flat $\log(A_i)$
\item Prior 2: flat $m_\chi$ and flat $A_i$
\item Prior 3: flat $\log(m_\chi)$ and flat $\log(A_i)$
\item Prior 4: flat $\log(m_\chi)$ and flat $A_i$
\end{itemize}
Fig.~\ref{fig:prior_compare} compares the resulting 95\% credibility contours on the coupling-coefficient amplitude and WIMP mass for each EFT operator calculated from the likelihoods  using each of the four prior choices.
For each operator, the 95\% contours calculated for each prior combination agree to within a factor of $\sim2.5$.

There are some sampling issues visible in Prior~3 and Prior~4.
The flat-log prior for the WIMP mass does not sample the high mass regions well.
When the flat amplitude prior, which does not sample the low amplitude regions well, is combined with the flat-log prior on the WIMP mass (Prior~4), this can cause entire regions along the 95\% contour to be poorly sampled.
This is most visible in Operator $\mathcal{O}_{10}$.
Both Prior~3 and Prior~4 show extra wiggles in the contour that are not visible in Priors~1 and 2.
When comparing the upper limits for all four prior combinations, the ability to sample the region of interest well becomes more important.
This is why Prior~1 was chosen as the best prior probability distribution combination for the main analysis presented in this article.
Because Priors~2 and Prior~3 performed similarly, the explicit choice of Prior~1 is not expected to introduce a notable bias towards improved sensitivity.

\begin{acknowledgements}
The SuperCDMS collaboration gratefully acknowledges technical assistance from the staff of the Soudan Underground Laboratory and the Minnesota Department of Natural Resources.
The CDMSlite detectors were fabricated in the Stanford Nanofabrication Facility, which is a member of the National Nanofabrication Infrastructure Network, sponsored and supported by the NSF.
Funding and support were received from the National Science Foundation, the U.S. Department of Energy (DOE), Fermilab URA Visiting Scholar Grant No. 15-S-33, NSERC Canada, the Canada First Excellence Research  Fund, the Arthur B. McDonald Institute (Canada),  the Department of Atomic Energy Government of India (DAE), the Department of Science and Technology (DST, India) and the DFG (Germany) -- Project No. 420484612 and under Germany's Excellence Strategy -- EXC 2121 ``Quantum Universe'' -- 390833306.
Femilab is operated by Fermi Research Alliance, LLC,  SLAC is operated by Stanford University, and PNNL is operated by the Battelle Memorial Institute for the U.S. Department of Energy under contracts DE-AC02-37407CH11359, DE-AC02-76SF00515, and DE-AC05-76RL01830, respectively.
\end{acknowledgements}

\bibliographystyle{apsrev4-1}
\bibliography{CDMSlite_Run2_EFT_PRD}

\end{document}